%% file: moot-paper.tex
\documentclass{sigplanconf}

\pdfoutput=1

\usepackage{etex}
\reserveinserts{28}

\usepackage{graphicx}
\usepackage{amsmath}
\usepackage{amsfonts}
\usepackage{amssymb}
\usepackage{tikz}
\usetikzlibrary{arrows}
\usepackage[all]{xy}
\usepackage{subfigure}
\usepackage{algorithm}
\usepackage{algorithmic}
\usepackage[thmmarks,standard]{ntheorem}

\title{Generic Programming of Reusable, High Performance Container
  Types using Automatic Type Hierarchy Inference and Bidirectional
  Antichain Typing}

\authorinfo{Wouter Kuijper}{W.Kuijper@utwente.nl}{University of
  Twente}
\authorinfo{Michael Weber}{M.Weber@utwente.nl}{University of Twente}

\authorpermission
\copyrightdata{February 16, 2011}

\newcommand{\concept}[1]{\emph{#1}}
\newcommand{\ms}[1]{\mathsf{#1}}
\newcommand{\mr}[1]{\mathrm{#1}}

\newcommand{\mc}[1]{\mathcal{#1}}
\newcommand{\mt}[1]{\text{\tt #1}}
\newcommand{\simulates}{\simeq}
\newcommand{\nada}{\varnothing}
\newcommand{\rem}[1]{}

\begin{document}

\maketitle

\abstract{
We introduce a new compile--time notion of type subsumption based on
type simulation.
We show how to apply this static subsumption relation to support a
more intuitive, object oriented approach to generic programming of
reusable, high performance container types.
As a first step towards an efficient implementation of the resulting
type system in a compiler we present a novel algorithm for
bidirectional type inference over arbitrary syntax graphs.
The algorithm uses the new static type subsumption relation to
compress the data that has to be stored for each node in the typeflow
graph.
During typeflow analysis this means that the set of types for a
given node can be symbolically represented using antichains instead of
using bitvectors or some other explicit set representation. 
This results in a typing algorithm that is both flexible and precise
and shows good performance on representative instances.
}

\section{Introduction}

Besides their useful role in enforcing partial correctness, types play
an important role in program synthesis. Not only does a well designed
type system prevent the programmer from specifying certain unsafe
operations, types also serve to disambiguate programs.  This is the
case for languages that support some form of \concept{function
  overloading} where argument types and return type determine the
particular function implementation that is invoked.

Many among the most popular programming languages to date are dynamic
languages. This means that, to a more or lesser degree, function
overloading is dealt with at run--time. This is not surprising as it
is generally more programmer--friendly than generic programming with
compile--time type--substitution. The latter technique constitutes the
only \emph{truly} static alternative that is available today for
writing high performance, reusable container types. The distinguishing
feature of the generic programming technique is that it makes use of
lexical substitution of types through \emph{type parameters}.

Using \emph{templates} and \emph{generic types} it becomes possible to
completely eliminate all overhead due to dynamic type checks because
all information about types can be fixed at compile--time. As such,
the aforementioned programming constructs are used mainly for
performance critical applications where the overhead of dynamic checks
to resolve overloading cannot be sustained.

However, obtaining this performance comes at a price. Programming
generic code can be difficult, labor intensive (due to the very
explicit way type parameters must be passed through every syntactical
construct) and counter intuitive. Quoting
Stroustrup~\cite{Stroustrup2000} on C++ templates:
\begin{quote}
  As far as the C++ language rules are concerned, there is no
  relationship between two classes generated from a single class
  template. For example:

  \verb+class Shape {/*...*/};+
  
  \verb+class Circle:public Shape {/*...*/};+

  Given these declarations, people sometimes try to treat a
  \verb+set<Circle*>+ as a \verb+set<Shape*>+. This is a serious
  logical error based on a flawed argument: ``A \verb+Circle+ is a
  \verb+Shape+, so a set of Circles is also a set of Shapes; [...]''

  \emph{Bjarne Stroustrup\\(The C++ Programming
    Language)}
\end{quote}

For most programmers this is counter intuitive: if a \verb+Circle+ is a
\verb+Shape+ then intuition tells us that a \verb+Set+ of
\verb+Circle+s must be a \verb+Set+ of \verb+Shape+s.
In Java Generics, the modern descendant of the C++ template system,
this particular situation has not improved. Quoting
Bracha~\cite{Bracha2004} on Java Generics:
\begin{quote}
  In general, if \verb+Foo+ is a subtype (subclass or subinterface) of
  \verb+Bar+, and G is some generic type declaration, it is not the
  case that \verb+G<Foo>+ is a subtype of \verb+G<Bar>+. This is
  probably the hardest thing you need to learn about generics, because
  it goes against our deeply held intuitions.

  \emph{Gilad Bracha\\(Generics in the Java Programming Language)}
\end{quote}

This counter intuitive trait in current generic programming approaches
is caused by the fact that type subsumption of the underlying
languages is essentially a dynamic notion that is designed to be
resolved at \emph{run--time}. So we see that there is a real need for
a programmer--friendly way of dealing with static type hierarchy
designed to be resolved at \emph{compile--time}.

As a solution to this problem we propose a new notion of \emph{static
  type subsumption}. This new subumption relation can be used in
conjunction with the currently prevalent notion of \emph{dynamic type
  subsumption}. By making this distinction more clearly we are
treating the static type subsumption relation as a first class
citizen. In particular this means that it becomes possible to use
generic programming \emph{in an object oriented style}, i.e.: if
\verb+Shape+ statically subsumes \verb+Circle+ this will imply that
\verb+Set<Shape>+ statically subsumes \verb+Set<Circle>+, and
\verb+List<Set<Shape>>+ statically subsumes \verb+List<Set<Circle>>+,
etc.\footnote{Given that {\tt List} and {\tt Set} are generic
  container types that support these type substitutions.}

The dynamic type subsumption relation will be, in general, a subset of
the static type subsumption relation where additional alignment
constraints must be met. In this paper we will focus exclusively on
computing and exploiting the static type subsumption relation.

\subsection{Contribution and Structure of the Paper}

The contribution of this paper is twofold. In Section~\ref{sec:types}
we present a type system that combines structural typing, function
overloading and static type strengthening as an intuitive, object
oriented alternative to existing generic programming approaches. In
Section~\ref{sec:typing} we present a new bidirectional antichain
typing algorithm that allows a practical, efficient implementation of
this new type system.

The paper is structured as follows.  In Section~\ref{sec:related} we
discuss related work. In Section~\ref{sec:motivating} we give a
motivating example which will also serve as a running example for
illustrating the definitions in the following sections. In
Section~\ref{sec:types} we give a formalization of the type hierarchy
as a simulation relation and we explain how to infer type hierarchy
from surface level declarations and definitions. In
Section~\ref{sec:typing} we show how to subsequently use the inferred
type hierarchy to efficiently type programs. In
Section~\ref{sec:conclusion} we give some perspectives on our current
results and future work.

\section{Related Work}\label{sec:related}

\concept{Types} were originally invented as simple names for sets of
values that form part of a programming language~\cite{Naur1963}. As
programming languages grew more sophisticated in keeping up with the
complexity of the problems they were employed to solve, types evolved
into more than simple sets of values and became an object of study in
and of itself. 

It was observed that types were instrumental in enforcing all kinds of
safety constraints thus obtaining a form of partial
correctness~\cite{Middleton1977,Cardelli1985}.

It was found independently by Hindley~\cite{Hindley1969} and
Milner~\cite{Milner1978} that polymorphic types were useful for
structuring programs.  Since the early contributions of Hindley and
Milner many alternatives and extensions to their approach have been
suggested in the literature~\cite{Schwartz1975,Schonberg1981,Palsberg1992,Aiken1993,Eifrig1995,Aiken1997}.

More recently other uses of types are being explored that enrich type
systems in various ways with constraints and qualifiers that can
express certain invariants and in this way further the role of types
in writing safe and correct programs~\cite{Foster2006,Nystrom2008}.

Another line of work proposes to improve the flexibility of typing
programs by treating the typing problem using a form of dataflow
analysis~\cite{Tenenbaum1974,Kaplan1980,Khedker2003}.
The work in this paper is related to the dataflow approach as we also
exploit the structure of the syntax graph to explicitly guide the
typing process. However we use an inherently bidirectional antichain
algorithm that does not immediately fit within the dataflow framework.

Antichains have recently received notable attention for their
potential use as a symbolic representation for upward or downward
closed sets. As such, with antichains we can solve many problems in
formal language theory much more efficiently than classical algorithms
that, typically, require a subset
construction~\cite{wulf_antichains:_2006}.

\section{A Motivating Example}\label{sec:motivating}

The {\sc moot} (modular object oriented template) programming layer is
a thin experimental programming layer over a small subset of C++. It
adds static function overloading and static type strengthening to the
basic C--like subset which forms the core of the C++ language.

Informally we say that type A \concept{statically subsumes} type B (or
type B is \concept{stronger} than type A) iff all the relevant
operations that can be compiled for type A can also be compiled for
type B. In effect this is a form of simulation relation between type A
and B with respect to the operations that objects of type A and B
support. We will discuss this formally in Section~\ref{sec:types}.

As \emph{relevant} operations (relevant to the type subsumption
ordering) we take all the built in operations, structural field select
operations, pointer and array dereference and the ``operation'' of
being passed to or returned from a function that is part of some
formal protocol definition.

As an example we consider the following {\sc moot}
definition of a protocol for iterating over a collection of values:
\begin{samepage}
\begin{verbatim}
protocoltype Iterable;
\end{verbatim}
\end{samepage}
\begin{samepage}
\begin{verbatim}
protocoltype Iterator;
\end{verbatim}
\end{samepage}
\begin{samepage}
\begin{verbatim}
void FIRST( Iterable c, Iterator &e );
\end{verbatim}
\end{samepage}
\begin{samepage}
\begin{verbatim}
bool DONE( Iterable c, Iterator e );
\end{verbatim}
\end{samepage}
\begin{samepage}
\begin{verbatim}
void NEXT( Iterable c, Iterator &e );
\end{verbatim}
\end{samepage}
\begin{samepage}
\begin{verbatim}
any DATA( Iterable c, Iterator e );
\end{verbatim}
\end{samepage}
As a first example we instantiate this protocol for counting up to
some integer value:
\begin{samepage}
\begin{verbatim}
void FIRST( int c, int &e ) { e = 1; }
\end{verbatim}
\end{samepage}
\begin{samepage}
\begin{verbatim}
bool DONE( int c, int e ) { return e > c; }
\end{verbatim}
\end{samepage}
\begin{samepage}
\begin{verbatim}
void NEXT( int c, int &e ) { e++; }
\end{verbatim}
\end{samepage}
\begin{samepage}
\begin{verbatim}
int DATA( int c, int e ) { return e; }
\end{verbatim}
\end{samepage}
Now the following is a valid application:
\begin{samepage}
\begin{verbatim}
int x = 5; int y;
for ( FIRST(x,y); !DONE(x,y); NEXT(x,y) ) {
  printf( "%d; ", DATA(x,y) );
}
\end{verbatim}
\end{samepage}
Which prints: \verb+1; 2; 3; 4; 5;+ Now let us instantiate this
protocol for integer intervals:
\begin{samepage}
\begin{verbatim}
struct _Ival {
  int min;
  int max;
};
\end{verbatim}
\end{samepage}
\begin{samepage}
\begin{verbatim}
typedef struct _Ival Ival;
\end{verbatim}
\end{samepage}
\begin{samepage}
\begin{verbatim}
void FIRST( Ival+ c, int &e ) { e = c.min; }
\end{verbatim}
\end{samepage}
\begin{samepage}
\begin{verbatim}
bool DONE( Ival+ c, int e ) { return e > c.max; }
\end{verbatim}
\end{samepage}
\begin{samepage}
\begin{verbatim}
void NEXT( Ival+ c, int &e ) { e++; }
\end{verbatim}
\end{samepage}
\begin{samepage}
\begin{verbatim}
int DATA( Ival+ c, int e ) { return e; }
\end{verbatim}
\end{samepage}

\begin{figure}\center
\input{figures/hierarchy_1.tex}
\caption{Inferred type subsumption order for some of the types from
  the Example in Section~\ref{sec:motivating}, the dashed areas show
  the downward closed sets of types denoted with {\tt Iterable+} and
  {\tt Ival+} respectively.}\label{fig:hierarchy:1}
\end{figure}
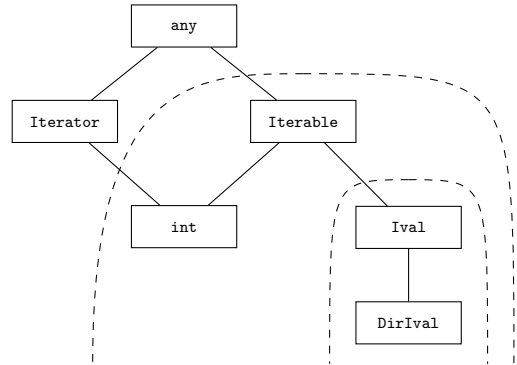

Here the plus \verb-+- type qualifier signals that the declared type
is \concept{compile--time strengthenable}, meaning the functions may
also be invoked with arguments of a stronger type taken from the
downward closed set of types under \verb-Ival+- in the subsumption
hierarchy of types as shown in Figure~\ref{fig:hierarchy:1}. We say the
type may be \emph{strengthened} if this is necessary to get a type
correct program.

Now that we have two different instantiations of the protocol it is
useful to have a generic function template for printing
\verb+Iterable+ collections:
\begin{samepage}
\begin{verbatim}
void print( Iterable+ c ) {
  Iterator+ e;
  for ( FIRST(c,e); !DONE(c,e); NEXT(c,e) ) {
    print( DATA(c,e), "; " );
  }
}
\end{verbatim}
\end{samepage}
\begin{samepage}
\begin{verbatim}
void print( any+ a, any+ b ) {
  print(a); print(b);
}
\end{verbatim}
\end{samepage}
\begin{samepage}
\begin{verbatim}
void print( int+ i ) {
  printf( "%d", i );
}
\end{verbatim}
\end{samepage}
\begin{samepage}
\begin{verbatim}
void print( char+ *s ) {
  printf( "%s", s );
}
\end{verbatim}
\end{samepage}
We use \verb+any+ as a special type that subsumes all types (in effect
it denotes the top element of the type lattice). The following is then
a valid application:
\begin{samepage}
\begin{verbatim}
Ival i; i.min = 11; i.max = 15;
print(i, "\n");
\end{verbatim}
\end{samepage}
Which prints: \verb+11; 12; 13; 14; 15;+
To give a slightly more interesting example, we want to subclass the
integer interval with a directed interval that includes an extra field
for iterating over the interval from either side using a given
increment. This can be done as follows:
\begin{samepage}
\begin{verbatim}
struct _DirIval {
  int min;
  int max;
  int delta;
};
\end{verbatim}
\end{samepage}
\begin{samepage}
\begin{verbatim}
typedef struct _DirIval DirIval;
\end{verbatim}
\end{samepage}
\begin{samepage}
\begin{verbatim}
void FIRST( DirIval+ c, int &e ) {
  if (c.delta > 0) e = c.min; else e = c.max;
}
\end{verbatim}
\end{samepage}
\begin{samepage}
\begin{verbatim}
bool DONE( DirIval+ c, int e ) {
  return (c.delta > 0 ? e > c.max : e < c.min);
}
\end{verbatim}
\end{samepage}
\begin{samepage}
\begin{verbatim}
void NEXT( DirIval+ c, int &e ) { 
  e += c.delta; 
}
\end{verbatim}
\end{samepage}
We may leave out the definition of \verb+DATA+ because it carries over
from the old definition with \verb+Ival+. The following is then a
valid application:
\begin{samepage}
\begin{verbatim}
DirIval d;  d.min = 11; d.max = 15; d.delta = -2;
print(d, "\n");
\end{verbatim}
\end{samepage}
Which prints: \verb+15; 13; 11;+.

\subsection{Translating Type Strengthening to C++}

Space constraints limit us from describing the {\sc moot} programming
layer in too much detail. However, since in the example we are using
only a single new syntactical construct (the \verb-+- strengthability
type qualifier) layered on top of the basic C language we will
explain the semantics of this construct by showing how it compiles
down to C++.
For example, the definitions of the protocol functions for the type
\verb+DirIval+ compile down to:
\begin{samepage}
\begin{verbatim}
void FIRST_DirIval_int( DirIval c, int &e ) {
  if (c.delta > 0) e = c.min; else e = c.max;
}
\end{verbatim}
\end{samepage}
\begin{samepage}
\begin{verbatim}
bool DONE_DirIval_int( DirIval c, int e ) {
  return (c.delta > 0 ? e > c.max : e < c.min);
}
\end{verbatim}
\end{samepage}
\begin{samepage}
\begin{verbatim}
void NEXT_DirIval_int( DirIval c, int &e ) { 
  e += c.delta; 
}
\end{verbatim}
\end{samepage}
\begin{samepage}
\begin{verbatim}
int DATA_DirIval_int( DirIval c, int e ) { 
  return e; 
}
\end{verbatim}
\end{samepage}
Note that the last function has been inherited from \verb+Ival+. Also
note that the type information has been pushed onto the identifier
names to make them unique. The print function for the same type
compiles down to:
\begin{samepage}
\begin{verbatim}
void print_DirIval( DirIval c ) {
  int e;
  for ( FIRST_DirIval_int(c,e); 
        !DONE_DirIval_int(c,e);
        NEXT_DirIval_int(c,e) ) {
    print_int( DATA_DirIval_int(c,e) );
  }
}
\end{verbatim}
\end{samepage}
As can be seen all the function overloading has been completely and
statically resolved by the {\sc moot} layer and the program can be
readily compiled by any C++ compiler.

Resolving calls to overloaded functions and strengthening the
strengthenable declarations to their proper types first requires us to
infer the type hierarchy in the form of the type subsumption relation
as shown in Figure~\ref{fig:hierarchy:1}. We come back to this in
Section~\ref{sec:types}.

\subsection{Container Types}\label{sec:container}

\begin{figure}\center
\input{figures/classdiag.tex}
\caption{Class--diagram for some of the types in the
  example.}\label{fig:classdiag}
\end{figure}
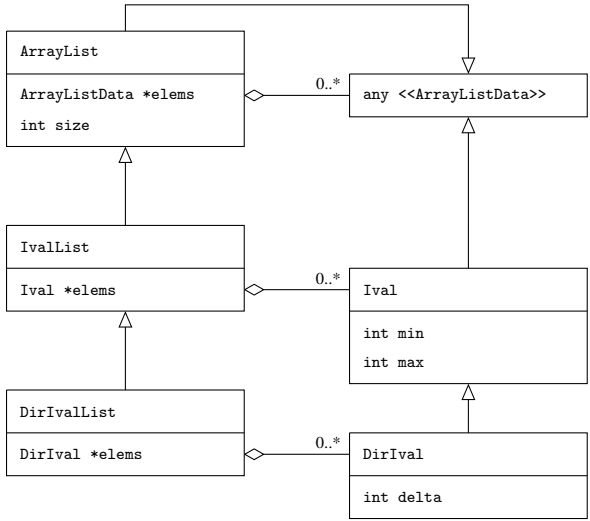

In the previous section we showed how the {\sc moot} layer is capable
of resolving function overloading and automatic type strengthening at
compile--time which are important prerequisites for an
object oriented generic programming layer. The second crucial
ingredient for any generic programming language is the instantiation
of types from generic container types.

Perhaps surprisingly it is not at all difficult to support the
instantiation of types from generic container types. This is mainly a
matter of lexical substitution of types which just requires the
appropriate syntax. The difficult part is to keep track of relations
between the resulting types \emph{after} these substitutions have been
carried out by the compiler.

Note in the previous example that we do not require the programmer to
make any explicit declaration of type subsumption. This is a very
important feature of the programming layer. It now becomes possible to
build container types whilst maintaining the static type subsumption
relation. For example, in {\sc moot} a simple, generic arraylist
structure might be defined as follows:
\begin{samepage}
\begin{verbatim}
parametertype ArrayListData;
\end{verbatim}
\end{samepage}
\begin{samepage}
\begin{verbatim}
struct _ArrayList {
  ArrayListData *elems;
  int size;
};
\end{verbatim}
\end{samepage}
\begin{samepage}
\begin{verbatim}
typedef struct _ArrayList ArrayList;
\end{verbatim}
\end{samepage}
This generic list datastructure can now be instantiated using
parameterized typedef declarations:
\begin{samepage}
\begin{verbatim}
typedef 
  ArrayList<Ival ArrayListData> 
  IvalList;
\end{verbatim}
\end{samepage}
\begin{samepage}
\begin{verbatim}
typedef 
  ArrayList<DirIval ArrayListData> 
  DirIvalList;
\end{verbatim}
\end{samepage}

In {\sc moot} it now follows automatically that \verb+IvalList+
statically subsumes \verb+DirIvalList+ without any additional help
from the programmer. So, in particular, it is possible to strengthen
any declaration of \verb-IvalList+- to \verb-DirIvalList+-.

The semantics of the \verb+<...>+ parameterized typedef construct can
be defined purely lexically: the type substitutions occurring between
the angled brackets are carried out on the original type definition
(and, transitively, on any \verb+struct+ or \verb+typedef+ definition
on which it depends).
For the example, the end result compiles down to normal type
declarations as follows:
\begin{samepage}
\begin{verbatim}
struct _IvalList {
  Ival *elems;
  int size;
};
\end{verbatim}
\end{samepage}
\begin{samepage}
\begin{verbatim}
typedef struct _IvalList IvalList;
\end{verbatim}
\end{samepage}
\begin{samepage}
\begin{verbatim}
struct _DirIvalList {
  DirIval *elems;
  int size;
};
\end{verbatim}
\end{samepage}
\begin{samepage}
\begin{verbatim}
typedef struct _DirIvalList DirIvalList;
\end{verbatim}
\end{samepage}
As can be seen the result depends on naming conventions: the original
type name is substituted with the new type name.

Note that, without a structural type system, it would be impossible to
conveniently maintain the proper type subsumption relation \emph{and},
at the same time, allow such a powerful lexical construct like the
\verb+<...>+ \verb+typedef+ parameter construct.

\subsection{Parameter Types with Protocol Assumptions}

Substitution of parameter types for arbitrary types is a common design
pattern used in generic programming due to the fact that container
types are usually intended to work for \emph{any} type. No assumptions
are made on the internal structure of the underlying data or on the
operations available for the underlying data.

However, opaque parameter types without any assumptions placed on them
are not always sufficient. There are certain cases where we would like
to provide specialized, or optimized functionality for data that
satisfies certain additional assumptions. One example would be a
datastructure for an ordered list that relies on a comparison function
being available over the underlying data. In {\sc moot} it possible to
formalize this additional assumption using protocols. As an example
consider the following refinement of our simple array list:
\begin{samepage}
\begin{verbatim}
protocoltype Comparable;
\end{verbatim}
\end{samepage}
\begin{samepage}
\begin{verbatim}
bool LTE( Comparable x, Comparable y );
\end{verbatim}
\end{samepage}
\begin{samepage}
\begin{verbatim}
parametertype SortedArrayListData : Comparable;
\end{verbatim}
\end{samepage}
\begin{samepage}
\begin{verbatim}
typedef 
  ArrayList<SortedArrayListData ArrayListData> 
  SortedArrayList;
\end{verbatim}
\end{samepage}

Now the new \verb+SortedArrayList+ parameter type inherits the
\verb+LTE+ (less--than--or--equal) protocol operation from protocol
type \verb+Comparable+. We use this new parameter type to define a
derived container type \verb+SortedArrayList+ that should keep the
elements in the arraylist sorted. We will not work this example out
further in this paper. However if we were to define the implementation
of \verb+SortedArrayList+ we would do so in terms of the parameter
type \verb+SortedArrayListData+. In each function that we would write
as part of this implementation we could then safely assume the
existence of a suitable function \verb+LTE+ that implements the
underlying ordering.

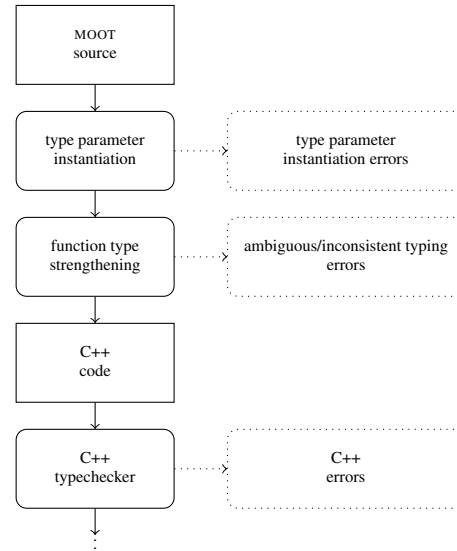
\begin{figure}[t]\center
\input{figures/early.tex}
\caption{Early warning of type errors in the case of {\sc moot}
  .}\label{fig:early}
\end{figure}

If the user would try to instantiate our new, sorted datatype without
defining a suitable \verb+LTE+ operation the compiler would detect
this by checking whether the instantiated type is subsumed by the
original declaration, i.e. if the user would now declare:
\begin{samepage}
\begin{verbatim}
typedef 
  SortedArrayList<Ival SortedArrayListData> 
  SortedIvalList;
\end{verbatim}
\end{samepage}
The compiler would give the following error message:
\begin{samepage}
\begin{verbatim}
SortedArrayListData does not subsume Ival
missing: LTE( (Ival), ... );
\end{verbatim}
\end{samepage}
\begin{figure}[t]\center
\input{figures/late.tex}
\caption{Reporting of type errors for C++ templates.}\label{fig:late}
\end{figure}
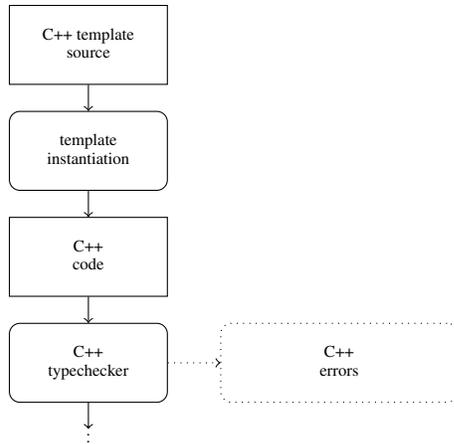
As shown in Figure~\ref{fig:late} the traditional approach to
template languages does not allow such checks to be performed before
the actual type--checking phase is entered. In Figure~\ref{fig:early}
we show how this situation is improved in {\sc moot}.

\subsection{Object Orientation and Type Hierarchy}

Because type hierarchy in {\sc moot} is inferred automatically, much
of the syntax that is traditionally present in object oriented
languages is missing in {\sc moot}\footnote{In this paper we do not
  discuss all the syntax that {\sc moot} offers. In particular it
  possible to explicitly inherit a {\tt struct} type from another {\tt
    struct} type. But in contrast to other object oriented languages
  this is syntactic sugar rather than a primitive
  construct.}. Nevertheless, the primitives we discussed so far offer
us enough freedom to build our own object systems conveniently. As
such {\sc moot} offers us the ability to use generic programming in an
object oriented style.

To illustrate this, Figure~\ref{fig:classdiag} shows some of the types
introduced in the running example. The diagram shows the proper type
subsumption relation as the inheritance relation and the type
relations induced by the struct fields as aggregations. The result is
a ``class--diagram'' of our types.

Note that, in order to keep the class--diagram compact, we left out
the operations (\verb+FIRST+, \verb+NEXT+, \verb+DONE+, \verb+DATA+,
etc.). These operations that take values of the various types as their
first arguments would typically be included in such a class--diagram
as \emph{methods}. In this context it is important to note that, in
order to infer this type hierarchy, we need to deal with a potential
form of circularity that arises when we adopt the proposed structural
definition of type subsumption together with the notion of type
strengthening for function arguments.

In fact this potential circularity \emph{is} present in the example.
In the one direction: the reason that \verb+DirIval+ is subsumed by
\verb+Ival+ is that all the field select operations (\verb+.min+,
\verb+.max+), and all the protocol operations (\verb+FIRST+,
\verb+NEXT+, \verb+DONE+, \verb+DATA+) which are available for
\verb+Ival+ are also defined for \verb+DirIval+. In the other
direction: the \verb+DATA+ operation is defined for \verb+DirIval+
because it is inherited from \verb+Ival+ and this only works precisely
because \verb+DirIval+ is subsumed by \verb+Ival+, which entails
\verb-Ival+- declarations may be \concept{strengthened} to
\verb-DirIval+- declarations.

In general the potential circularity of reasoning can be broken by
defining the type subsumption relation as the largest possible type
simulation relation that is mutually consistent with the rules for
\verb+protocoltype+ and \verb+struct+ subsumption. We will come back
to this in Section~\ref{sec:types}.

\subsection{Calling Functions with Strengthenable Arguments}

So far we have discussed how parameter types allow us to parameterize
and instantiate generic type declarations. In order to implement these
types we need to define functions over them. In {\sc moot} function
definitions are never instantiated through the \verb+<...>+ angled
bracket notation, this notation is reserved for type declarations. For
function definitions we rely solely on \emph{type strengthening}. For
this it is important to understand how a call to a function with
strengthenable arguments is resolved, i.e.: which of the various
overloaded function bodies is actually invoked?  As an example we take
the previously defined \verb+print/1+ function.

In the example we overloaded the \verb+print/1+ function several
times. Now we did not define \verb+print/1+ as part of a formal
protocol. As such it is does not influence the type subsumption
relation. However, in the other direction, the type subsumption
relation \emph{does} influence how calls to \verb+print/1+ get
resolved.

In particular we gave a definition with signature \verb-print(int+)-
for simple integers, and we gave a second definition with signature
\verb-print(Iterable+)- for \verb+Iterable+ values. At the same time
we implemented the \verb+Iterable+ protocol for simple integers. This
means that we need a principle on which to resolve a call like:
\verb+print(3)+: do we map it to the former or to the latter function
definition?

We answer this following the usual semantics which means we resolve to
the \emph{strongest possible} signature. In this case the signature
\verb-print(int+)- is stronger than the signature
\verb-print(Iterable+)- because \verb-int- is \verb-Iterable- but not
the other way around. In {\sc moot} we provide syntax to fine tune the
matching of the function on one or more arguments by weakening the
signature against which the function is matched. For the example we
might write \verb+print( [^Iterable]3 )+. Which prints: \verb+1; 2; 3;+

\subsection{Multiple Strengthenable Arguments}

The only remaining issue concerning the semantics of the new
strengthenable type qualifier arises when there is \emph{more than
  one} strengthenable formal parameter to some defined function.

In order to understand what would be the right call matching semantics
for functions with multiple strengthenable arguments it is good to look at
some pathological cases and see how we should best deal with these
cases, that is: providing minimal confusion to the programmer. First,
consider the situation where we would provide the following two
function definitions with the same name and arity:
\begin{samepage}
\begin{verbatim}
DirIval+ intersect( DirIval+ i1, Ival+ i2 )     (1)
\end{verbatim}
\end{samepage}
\begin{samepage}
\begin{verbatim}
DirIval+ intersect( Ival+ i1, DirIval+ i2 )     (2)
\end{verbatim}
\end{samepage}
We say these two function signatures are \emph{incomparable}, because
the first definition is stronger in the first argument type, whereas
the second definition is stronger in the second argument type. When we
consider which of the functions to call in an application like the
following:
\begin{samepage}
\begin{verbatim}
DirIval d1, d2, d3;
...
d1 = intersect( d2, d3 );
\end{verbatim}
\end{samepage}
It follows we must either pick one of the two function definitions or
we must reject the program with a typing error. The first option is
problematic because it introduces an element of arbitrariness into the
semantics. Therefore, in this case, we prefer the second option.

\subsection{Singleton Antichain Semantics} 

Now consider what should happen if, in addition to function
definitions (1) and (2) we would add a third function definition:
\begin{samepage}
\begin{verbatim}
Ival+ intersect( Ival+ i1, Ival+ i2 )           (3)
\end{verbatim}
\end{samepage}
For our example, with respect to the call matching semantics for
\verb+intersect(d2, d3)+, we have a third option to consider namely to
invoke function definition (3). Even though it is strictly
\emph{weaker} than function definitions (1) and (2), it at least lacks
the element of arbitrariness. To see this just note that the antichain
of incomparable functions with the same name and arity as function
definition (3) contains only function definition (3) itself. As such,
we will refer to this tentative semantics as the \emph{singleton
  antichain semantics}.

The singleton antichain semantics does not suffer from the
arbitrariness we discussed earlier. However, there is another reason
to reject this semantics. In practice what happens when programmers
use function overloading is that definitions are grouped,
conceptually, into \emph{classes} which often also end up being
defined in different source files (\emph{modules}). For our example
this might mean that function definitions (1), (2) and (3) occur far
removed from each other.

Now consider a programmer who completes function definition (3)
\emph{first} and subsequently goes on and overloads this definition
with function definition (2). The program might compile and work for a
while until, at some point, somebody decides to add function
definition (1) without paying attention to the existence of function
definition (2). Given the singleton antichain semantics this would
mean that the new function definition (1) would be silently ignored
because of the existence of function definition (2), and, vice versa,
the existing function definition (2) would now also be silently
ignored because of the existence of the new function definition
(1). More seriously even, in all the cases where we used to invoke
function definition (2) we would then \emph{go back} to the invocation
of the \emph{older} function definition (3). So we see that adding a
new function under such a call matching semantics may have unexpected,
non--local effects.

\subsection{Strongest Call Semantics}\label{sec:strongestcall}

The latter example shows that the singleton antichain call matching
semantics would also be problematic. For this reason we propose the
\emph{strongest call semantics}. Under this call matching semantics a
function definition is only invoked iff \emph{all} the formal
parameters in the signature, pointwise, are the strongest possible
match to the types of the corresponding actual parameters among all of
the defined function signatures (of the same name and arity). Under
this semantics the call \verb+intersection(d2, d3)+ remains untypeable
also when function definition (3) is added. To fix this situation the
programmer may always introduce a fourth function definition:
\begin{samepage}
\begin{verbatim}
DirIval+ intersect( DirIval+ i1, DirIval+ i2 )  (4)
\end{verbatim}
\end{samepage}
To avoid the non--local effects that we mentioned earlier we may warn
the user when two incomparable function definitions (with the same
name and arity) occur in different source files. Together with the
strongest call semantics this enforces a reasonable level of
modularity.
In Section~\ref{sec:typing} we will show how to formalize and enforce
the strongest call semantics for function call expressions.

\section{Inferring Type Hierarchy}\label{sec:types}

In Section~\ref{sec:motivating} we already briefly remarked that we
view type subsumption as a form of simulation relation, in this
section we will discuss this formally. The analysis in this section is
based on the assumption that we can obtain a finite set of relevant
types from the source code of the program. We will assume that the
user will provide all the relevant types\footnote{In practice (for
  {\sc moot}) we do saturate the set of relevant types with 1--deep
  pointer types to avoid overly pedantic errors, further note that
  instantiating a parameterized container type may introduce a
  significant amount of types without any work for the user, given
  that any type that the container type transitively depends on is
  also instantiated automatically.}. If it turns out that the program
cannot be typed because the user forgot to provide a type this will be
flagged with a clear error message. However, the typing procedure will
never introduce new types outside the finite set of relevant types. In
this way we ensure termination. Below we give the basic definition of
type subsumption as a simulation relation.

\begin{definition}[Type Simulation and Subsumption]
Let $T$ be a finite set of \concept{relevant
  types}, and let $\{ R_\sigma \}_{\sigma \in \Sigma}$ be a finite,
indexed set of \concept{type relations} over $T$, i.e. for all $\sigma
\in \Sigma$ it holds $R_\sigma \subseteq T \times T$. For all $\sigma
\in \Sigma$ let $R^\dagger_\sigma \subseteq R_\sigma$ be a selected
subset of the type relation, we say the tuples in $R^\dagger_\sigma$
are \concept{strengthenable}.
For a given candidate subsumption relation $\mc{R}_\preceq \subseteq T
\times T$ and any two types $t, t' \in T$ we define $t
\simulates_{\mc{R}_\preceq}^\sigma t'$ iff for all $u' \in T$ such that $t'
R_\sigma u'$ there exists some $u \in T$ such that $u \mc{R}_\preceq u'$
and $t R_\sigma u$ or $t (\mc{R}_\preceq ; R_\sigma^\dagger) u$ where $;$
denotes relation composition.
We say some candidate subsumption relation $\mc{R}_\preceq \subseteq T
\times T$ is a $\sigma$--simulation relation iff for all $(t, t') \in
\mc{R}_\preceq$ it holds that $t \simulates_{\mc{R}_\preceq}^\sigma t'$.
We say some candidate subsumption relation $\mc{R}_\preceq \subseteq T
\times T$ is \concept{valid} iff for all $\sigma \in \Sigma$ it holds
$\mc{R}_\preceq$ is a $\sigma$--simulation.
To compute the greatest valid subsumption relation based on this
requirement we may define the following fixed point operation:
\begin{align*}
F( \mc{R}_\preceq ) &= \bigcap_{\sigma \in \Sigma} \{ (t, t') \in
\mc{R}_\preceq\ |\ t \simulates_{\mc{R}_\preceq}^{\sigma} t' \}
\end{align*}
Since $F(\cdot)$ is monotone it always has a greatest fixed point. Let
$\mc{R}^\ms{init}_\preceq$ be some pre--order that forms an initial,
syntactical overapproximation of the desired type subsumption
relation. We now define $\preceq$ as the \emph{largest valid
  subsumption relation} contained in $\mc{R}^\ms{init}_\preceq$.
When we speak about the \concept{simulation graph} we mean the graph
over $T$ that includes $\preceq$ and all the other type relations $\{
R_\sigma \}_{\sigma \in \Sigma}$.
\end{definition}

This definition is generally applicable and still allows a great deal
of freedom in the actual details of the type system. We will now
sketch a number of examples where we make use of the definition of
type subsumption as a simulation relation in the {\sc moot} type
system.

\subsection{The Role of Type Hierarchy in {\sc moot}}

The {\sc moot} programming layer is specifically meant to deal with
\emph{static type subsumption}. In a performance oriented language
like C++ this is an important concept: if everything about types is
known at compile--time this means we can avoid introducing run--time
checks, avoid tagging data--structures with type--identifiers, and
employ the C++ compiler to optimize many operations.

In this context it is important to note that the structural type
system is introduced mainly to free the programmer from the burden of
having to deal with an overly rigid type system, and to do so with
minimal impact on code readability and efficiency.
As such the type subsumption relation that we will infer using the
techniques described here should \emph{not} be thought of as the final
safeguard that stands between the program and its execution.
Instead, the type subsumption relation serves mainly as an aid to
structure the set of types which benefits both the user (by allowing
concise and understandable code) as well as the typing procedure (by
allowing the partial order structure to be exploited for efficiency).

The final result will be compiled down to statically typed C++ and any
missing or illegal operations that are not caught by the {\sc moot}
type system will be rejected at the next stage of compilation, since
the C++ type system is more strict than the {\sc moot} type system.
The syntactical layer is thin enough to allow the user to understand
how an error message from the C++ compiler relates back to the
original code. This is especially true because line--numbering,
statements, expressions and control flow are fully preserved.
Also, because we do not present the user with a raw trace of the type
algebraic expressions (as is done in a template language like STL) the
error messages are in fact more understandable in the case of {\sc
  moot}.

\subsection{Simulation of Structural Types}

\begin{figure}\center
\input{figures/hierarchy_2.tex}
\caption{Inferred type subsumption order for some of the function
  types of the Example in Section~\ref{sec:motivating}, the dashed
  area shows the downward closed set of types denoted by the antichain
  of function types: $\{ \mt{int(*)(int,int)}, \mt{int(*)(Ival+,
    int)}\}$.}\label{fig:hierarchy:2}
\end{figure}
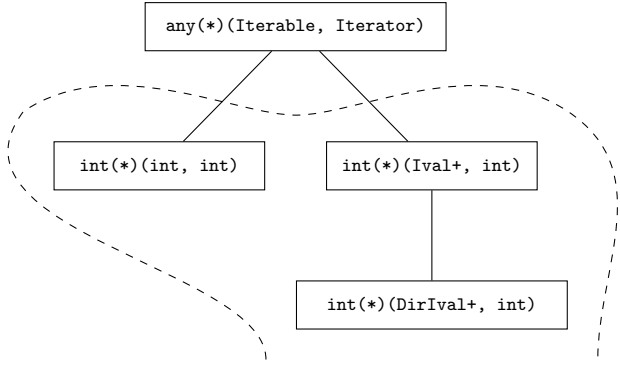

As a first example we will consider the structural types \verb+Ival+
and \verb+DirIval+ as they were defined in
Section~\ref{sec:motivating}, and the $\mt{.min}$ field selection
operation. We will formalize this as a type relation $R_\mt{.min}$
such that $(t, u) \in R_\mt{.min}$ iff $t$ is a \verb+struct+ type
that contains a field \verb+min+ of type $u$. For the example this
becomes:
\begin{align*}
R_\mt{.min} = \{ \mt{Ival} \mapsto \mt{int},\ \mt{DirIval} \mapsto
\mt{int} \}
\end{align*}
Now in order to see that \verb+DirIval+ simulates \verb+Ival+ with
respect to this type relation we must check whether the result after
applying \verb+.min+ on \verb+DirIval+ still simulates the result
after applying the same operation on \verb+Ival+. This simulation
condition can be summarized in the following subdiagram of the simulation graph:
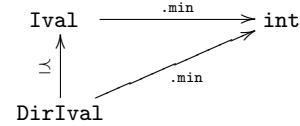
\begin{align*}
\xymatrix{
  \mt{Ival }\ar@{<-}[d]_{\preceq} \ar@{->}[rr]^-{\mt{.min}} 
  & & \mt{int} \\ 
  \mt{DirIval} \ar[rru]_-{\mt{.min}} 
}
\end{align*}
In this case the simulation requirement is fulfilled. For simple
direct type relations like selecting a field in a \verb+struct+ it is
quite straightforward to check the simulation requirement. This
approach also works for recursive \verb+struct+ types.

\subsection{Simulation of Function Types}

Treating function types is somewhat more involved.We give as an
example the \verb+DATA+ operation. On the level of types we formalize
the type relation $R_{\mt{DATA}_{1/2}}$ meaning ''possible signature
based on first argument type to the binary operation \verb+DATA+''.
More specifically, we place a pair $(t, f)$ of a type and a function
type in the argument--signature relation $(t, f) \in
R_{\mt{DATA}_{1/2}}$ iff there exists a \emph{definition} of binary
operation \verb+DATA+ with \emph{declared signature} $f$ that contains
$t$ as the first argument, moreover we put $(t, f) \in
R^\dagger_{\mt{DATA}_{1/2}}$ if, in addition, $t$ is marked as a
strengthenable argument in $f$. For the example this becomes:
\begin{align*}
 R_{\mt{DATA}_{1/2}} &= \\
 \{\ &\mt{Iterable} \mapsto \mt{any(*)(Iterable, Iterator)},\ \\
  &\mt{int} \mapsto \mt{int(*)(int, int)},\ \\
  &\mt{Ival} \mapsto \mt{int(*)(Ival+, int)}\ \} \\
 R^\dagger_{\mt{DATA}_{1/2}} &= \\
  \{\  &\mt{Ival} \mapsto \mt{int(*)(Ival+, int)}\ \}
\end{align*}
Here \verb-any(*)(Iterable, Iterator)- is the C type signature
notation for a function that takes a value of \verb+Iterable+ type as
its first argument, a value of \verb+Iterator+ type as its second
argument and returns a value of \verb+any+ type.

The simulation requirement between \verb+Ival+ and the \verb+Iterable+
protocol, on the first argument of the binary \verb+DATA+ operation,
can then be summarized in the following subdiagram of the simulation
graph:
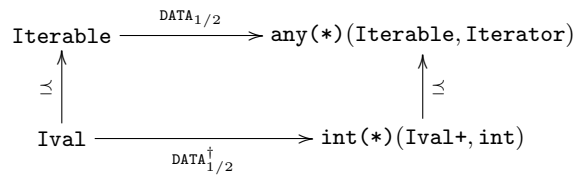
\begin{align*}
\xymatrix{
  \mt{Iterable} \ar@{<-}[d]_{\preceq} \ar@{->}[rr]^-{\mt{DATA}_{1/2}} 
  & & \mt{any(*)}(\mt{Iterable}, \mt{Iterator}) \ar@{<-}[d]^{\preceq} \\
  \mt{Ival} \ar[rr]_-{\mt{DATA}^\dagger_{1/2}} 
  & & \mt{int(*)}(\mt{Ival+}, \mt{int})
}
\end{align*}
We see that this simulation requirement depends on the simulation
between two function types. So we need to be explicit about when two
function types, like the ones shown in the diagram, are in the
subsumption relation. To do this within the current framework we
introduce new type relations $\mt{arg}_{i/j}$ for selecting the
$i$--th argument type from a $j$--ary function type:
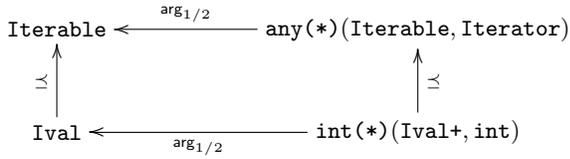
\begin{align*}
\xymatrix{
  \mt{Iterable} \ar@{<-}[d]_{\preceq} \ar@{<-}[rr]^-{\ms{arg}_{1/2}} 
  & & \mt{any(*)}(\mt{Iterable}, \mt{Iterator}) \ar@{<-}[d]^{\preceq} \\
  \mt{Ival} \ar@{<-}[rr]_-{\ms{arg}_{1/2}} 
  & & \mt{int(*)}(\mt{Ival+}, \mt{int})
}
\end{align*}
So in effect we see that the simulation requirement runs in both
directions in this case. For the second argument this works
likewise. For the return type we introduce a type relation
$\mt{ret}_{/j}$ for selecting the return type from from a $j$--ary
function type:
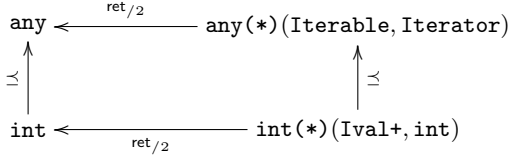
\begin{align*}
\xymatrix{
  \mt{any} \ar@{<-}[d]_{\preceq} \ar@{<-}[rr]^-{\ms{ret}_{/2}} 
  & & \mt{any(*)}(\mt{Iterable}, \mt{Iterator}) \ar@{<-}[d]^{\preceq} \\
  \mt{int} \ar@{<-}[rr]_-{\ms{ret}_{/2}} 
  & & \mt{int(*)}(\mt{Ival+}, \mt{int})
}
\end{align*}
More precisely; a function type A subsumes a function type B iff they
are of the same arity and all the arguments of A subsume the
corresponding arguments of B and the return type of A subsumes the
return type of B and at least all the places in which A is
strengthenable are also strengthenable in B (the latter condition is
enforced through $\mc{R}^\ms{init}_\preceq$). 

As can be seen the subsumption condition on function types is
covariant between arguments and return type. In this context it is
good to recall that we are considering only \emph{static type
  subsumption}.  For \emph{dynamic type subsumption} one might expect
a contravariant condition here.

In a \emph{dynamic} setting functions may get passed around.  The
relevant question is: can function A be called in a all contexts where
function B can be called? As such when building a dynamic subsumption
relation we should treat argument types as assumptions on the calling
context and return types as guarantees to the calling context. This
would lead to a contravariant definition.

However, in our \emph{static} setting, functions do not get passed
around in the same way. The relevant question is: can function A be
strengthened to all the signatures to which function B can be
strengthened? Static type subsumption (at compile time) is used for a
completely different programming intend, as such it should be
distinguished from, and possibly used in conjunction with, dynamic
type subsumption (at run time).

\subsection{Simulation of Strengthenable Function Types}

Next we consider how the strengthenable subset of the
argument--signature relation $\mt{DATA}^\dagger_{1/2}$ interplays with
the subsumption relation $\preceq$. For this we consider the example
of invoking the \verb+DATA+ operation on an object of type
\verb+DirIval+.

In particular we note that \verb-int(*)(DirIval+, int)- is \emph{not}
reachable from \verb+DirIval+ through the $R_{\mt{DATA}_{1/2}}$
argument--signature relation because we have not overloaded
\verb+DATA+ to that signature. Instead, we relied on argument
strengthening to carry over the definition with type signature
\verb-int(*)(Ival+, int)-. This is summarized in the following
subdiagram of the simulation graph:
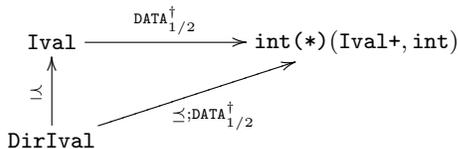
\begin{align*}
\xymatrix{
  \mt{Ival} \ar@{<-}[d]_{\preceq} \ar@{->}[rr]^-{\mt{DATA}^\dagger_{1/2}} 
  & & \mt{int(*)}(\mt{Ival+}, \mt{int}) \\
  \mt{DirIval} \ar@{->}[rru]_-{\preceq ; \mt{DATA}^\dagger_{1/2}} 
}
\end{align*}
In effect the presence of the \verb-+---qualifier prompts us to
transitively close the $R_{\mt{DATA}_{1/2}}^\dagger$ relation over the
$\preceq$ relation when selecting a function signature. If we had
omitted the \verb-+- from the definition of \verb+DATA+ this would
mean the first, horizontal arrow would not have been marked with
$\dagger$ and hence the second, diagonal arrow would not be present in
the diagram, which in turn would mean that this simulation requirement
would be violated and \verb+DirIval+ would no longer be subsumed by
\verb+Ival+:
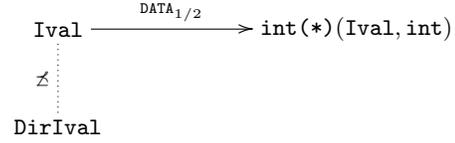
\begin{align*}
\xymatrix{
  \mt{Ival} \ar@{..}[d]_{\not\preceq} \ar@{->}[rr]^-{\mt{DATA}_{1/2}} 
  & & \mt{int(*)}(\mt{Ival}, \mt{int}) \\
  \mt{DirIval}
}
\end{align*}

\subsection{Simulation of Pointer Types}

Finally we consider how to deal with pointer types. We look at the
type subsumption between \verb+IvalList+ and \verb+DirIvalList+ as
introduced in Section~\ref{sec:container}. We see that these types are
defined in terms of the types \verb+Ival*+ and \verb+DirIval*+. To
handle such pointer types we introduce a new type relation $R_*$ such
that $(t_*, t) \in R_*$ iff $t_*$ is a pointer--type to type $t$. This
then gives the following subdiagram:
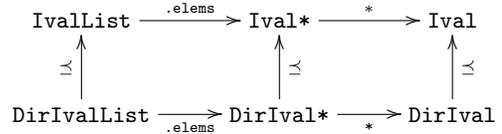
\begin{align*}
\xymatrix{
  \mt{IvalList} \ar@{<-}[d]_{\preceq} \ar@{->}[r]^-{\mt{.elems}} 
  & \mt{Ival*} \ar@{<-}[d]^{\preceq} \ar@{->}[r]^-{\mr{*}}
  & \mt{Ival} \ar@{<-}[d]^{\preceq} \\
  \mt{DirIvalList} \ar[r]_-{\mt{.elems}} 
  & \mt{DirIval*} \ar[r]_-{\mt{*}}
  & \mt{DirIval}
}
\end{align*}
which shows how pointer type subsumption is reduced to type
subsumption of the pointed--to types.

\subsection{Implementation Issues}

We already gave an abstract algorithm for
computing the largest valid subsumption relation by iterating a basic
fixed point operation. 
There are several important refinements that can be made to the basic
fixed point iteration to make it more efficient. 

The first is the computation of $\mc{R}^\ms{init}_\preceq$, the syntactical
overapproximation of $\preceq$. It is possible to eliminate many edges
in the simulation graph before starting the algorithm proper, just by
looking at the surface syntax of the declarations. As an example we
mention the comparison of \verb+struct+s based on the fields they have
available: if a \verb+struct+ A misses a field that is present in
\verb+struct+ B than B for sure does not subsume A. In our current
implementation we use a forward definition of syntactical subsumption
that uses a bounded depth exploration of \verb+struct+s and pointer
structures to compare two types for initial subsumption.

In this context it is important to note that we bound the number of
aforementioned comparisons by exploiting the pre--order structure
using appropriate datastructures. This means a new type can be
introduced into the pre--order by traversing the pre--order antichains
layer by layer gradually narrowing in on the set of direct parents of
the new type. This significantly reduces the number of comparisons
that need to be carried out.

A second optimization we apply is to use an edge elimination
algorithm based on a waitinglist for the closure procedure rather than
a naive fixed point iteration. This approach scales better because it
does not require a complete recompute of the simulation relation at
each iteration as the naive fixed point computation would.

Upon termination $\mc{R}_\preceq$ will be a \emph{pre--order}. The
typing algorithm we will describe in Section~\ref{sec:typing} expects
a \emph{partial order} on the set of types. Classically this is solved
by moving to the set of equivalence classes of types rather than
individual types, \emph{or}, alternatively, to mark all the equivalent
types as distinct.

For flexibility we give the user the choice to normalize types that
share an equivalence class to a single representant \emph{after}
hierarchy has been elicited, or, alternatively, to mark two or more
types as distinct \emph{before} hierarchy is elicited. Because it is
not really central to the current exposition, in the rest of the paper
we will assume that $\preceq$ is already a partial order.

\subsection{Error Reporting}

Whenever the user places an explicit request
for type subsumption the compiler will check whether this request is
congruent with the inferred type subsumption relation. If it turns out
that this is not the case it is important to provide the user with a
clear counterexample as to \emph{why} this is not the case. 

It is possible to do this by presenting the user, conceptually, with a
path through the simulation graph. This path will lead \emph{from} the
types that were requested to be in subsumption relation (but were not
in actuality), \emph{over} one or more type relation edges, \emph{to}
a point in the simulation graph were a clear contradiction is reached.

A contradiction usually means that the requested subsuming type
supports an operation that the requested subsumed type does not. Such
a path through the type simulation graph can subsequently be turned
into a normal C expression, indicating where an expression of the
requested subsumed type should be inserted to reach a contradiction
and using ellipsis for open terms that are not immediately relevant to
the counterexample. As an example of this consider the situation where
we request \verb+Ival+ to be subsumed by \verb+Iterator+, we can do
this using the following syntax:
\begin{samepage}
\begin{verbatim}
<check Ival subsumed by Iterator>
\end{verbatim}
\end{samepage}
Note that we are not \emph{changing} the subsumption relation in this
way. That would not be possible because the subsumption relation is
defined mathematically based on the operations that we provide. We are
not providing a new operation here, instead, we are merely stating a
property about our program which we want to maintain.  In this case
the property does not hold and the compiler will report the following
error:
\begin{samepage}
\begin{verbatim}
Iterator does not subsume Ival
missing: FIRST( ..., (Ival) );
\end{verbatim}
\end{samepage}
Clearly this gives us enough information to understand why the
subsumption does not hold, and also it gives us a starting point if we
wanted to fix this situation.

Using rules that turn each type relation that could potentially lead
to a contradiction into such a C-like expression term the compiler can
output similarly intuitive error messages for pointer types,
\verb+struct+ types, function types, etc.

\section{Bidirectional Antichain Typing}\label{sec:typing}

In this section we present the bidirectional antichain typing
algorithm.
First, we introduce our definition of \concept{syntax graph}.
Next, we introduce some auxiliary definitions concerning
\concept{antichains of types}. 
Finally, we introduce the bidirectional antichain typing algorithm
proper.

\begin{definition}[Syntax Graphs]\label{def:syntax_graphs}
Let $T$ be a set of \emph{relevant types} as before. \emph{A type
  relation} $M \subseteq T \times T$ is \emph{cross--closed} iff for
any pair of crossing edges $(t, u'), (t', u) \in M$ such that $t'
\preceq t$ and $u' \preceq u$ it holds $(t, u) \in M$. Let
$\{M_\gamma\}_{\gamma \in \Gamma}$ be an indexed set of cross--closed
\emph{type matching relations}.
A \concept{syntax graph} $G$ is
defined as a tuple $G = \langle N, E, \gamma \rangle$ where $N$ is a
set of \concept{nodes}, $E \subseteq N \times N$ is a set of
\concept{edges}, $\gamma : E \to \Gamma$ is a mapping assigning each
syntax edge a type matching relation. We define a \concept{simple
  typing} $\delta : N \to T$ as a map that assigns each syntax node a
type. We say $\delta$ is \emph{valid} iff for all edges $(n, m) \in E$
it holds $(\delta(n), \delta(m)) \in M_{\gamma(n, m)}$.
\end{definition}

Note that cross--closedness is a symmetric condition: a relation is
cross--closed iff its inverse is cross--closed. Further note that if a
relation is monotone or antitone it is trivially cross--closed.

A syntax graph is a structure that can be seen as a straightforward
generalization of a \concept{syntax tree} where we allow free--form
dependencies that transcend the basic syntax tree form. In practice
the syntax graph is built over the syntax tree after resolving
identifier/declaration dependencies.

The general definition of a syntax graph and type matching relations
still allows a lot of freedom in the actual formalization of the
particular type system we are interested in. Below we give two
particular examples of how this definition is used to encode the
typing rules for the {\sc moot} programming layer.

\subsection{Typing Struct Field Select Expressions}

First we give a basic example of a syntax graph. Consider
Figure~\ref{fig:typing:fwd:a}. This syntax graph corresponds to the
field select expression \verb+i.min+ as it was used in the example in
Section~\ref{sec:motivating}. It consists of two nodes. Node 1
represents the type of the \verb+i+ identifier and node 2 represents
the type of the \verb+i.min+ field. The only edge is labeled with the
type matching relation $\mt{.min}$. We will define this type matching
relation as:
\begin{align*}
M_\mt{.min} = \{ \mt{Ival} \mapsto \mt{int},\ \mt{DirIval} \mapsto
\mt{int} \}
\end{align*}
This type matching relation can be directly transferred from the type
simulation graph of Section~\ref{sec:types}, i.e.: $M_\mt{.min} =
R_\mt{.min}$.
In Figures~\ref{fig:typing:fwd:b} and~\ref{fig:typing:fwd:c} we show two
examples of a simple typing that assigns each node a type. Note that
only the typing shown in Figure~\ref{fig:typing:fwd:c} is valid.

\subsection{Typing Function Call Expressions}

As a second example we will show how function call expressions can be
represented and typed. Consider Figure~\ref{fig:typing:fun:a}. This
syntax graph corresponds to the function call expression
$\verb+DATA(x,y)+$ as it was used in the example in
Section~\ref{sec:motivating}. It consists of four nodes. Node 1
represents the type of the first argument, node 2 represents the type
of the second argument, node 3 represents the declared signature type
of the function that is being called, and finally node 4 represents
the type of the result that is returned by the function.

As can be seen the syntax graph in Figure~\ref{fig:typing:fun:a} is
decorated with three different type matching relations.
One for the first argument type, one for the second argument type and
one for the return type.  These three relations together determine the
call matching semantics. From Section~\ref{sec:strongestcall} we
recall the notion of \emph{strongest call semantics}. According to
this semantics we may only invoke a function if each of the argument
types in its declared signature is the best possible match to the
actual arguments that are provided. It is possible to enforce this by
defining the type matching relation accordingly. For the example this
becomes:
\begin{align*}
 M_{\mt{DATA}^\ms{arg}_{1/2}} &= \\
 \{\ &\mt{Iterable} \mapsto \mt{any(*)(Iterable, Iterator)},\ \\
  &\mt{int} \mapsto \mt{int(*)(int, int)},\ \\
  &\mt{Ival} \mapsto \mt{int(*)(Ival, int)},\ \\
  &\mt{DirIval} \mapsto \mt{int(*)(Ival, int)}\ \} \\
 M_{\mt{DATA}^\ms{arg}_{2/2}} &= \\
 \{\ &\mt{Iterator} \mapsto \mt{any(*)(Iterable, Iterator)},\ \\
  &\mt{int} \mapsto \mt{int(*)(int, int)},\ \\
  &\mt{int} \mapsto \mt{int(*)(Ival, int)}\ \} \\
 M_{\mt{DATA}^\ms{ret}_{/2}} &= \\
 \{\ &\mt{any} \mapsto \mt{any(*)(Iterable, Iterator)},\ \\
  &\mt{int} \mapsto \mt{int(*)(int, int)},\ \\
  &\mt{int} \mapsto \mt{int(*)(Ival, int)}\ \}
\end{align*}
These relations can easily be computed from the type simulation
graph. For example the first type matching relation
$M_{\mt{DATA}^\ms{arg}_{1/2}}$ can be computed in terms of the type
relations we introduced in Section~\ref{sec:types}:
\begin{align*}
M_{\mt{DATA}^\ms{arg}_{1/2}} &= R_{\mt{DATA}_{1/2}} \cup ( \preceq ;
R^\dagger_{\mt{DATA}_{1/2}} )
\end{align*}
Note that, in general, we must take care to remove from the resulting
relation any argument--signature pairs that do not satisfy the
strongest call semantics. For the example there are no such pairs. 
Note that the given relations are monotone by virtue of the type
simulation requirement, hence they are also, trivially, cross--closed.

In Figures~\ref{fig:typing:fun:b},~\ref{fig:typing:fun:d}
and~\ref{fig:typing:fun:e} we show several examples of a simple typing
that assigns each node a type. Note that only~\ref{fig:typing:fun:e} is a
valid typing. Figure~\ref{fig:typing:fun:c} is \emph{not} a simple
typing, as can be seen node 3 receives two different types, we will
show how to deal with such ambiguous typings in
Section~\ref{sec:antichains}.

\subsection{Type Promotion}

Since all the type relations that we gave so far satisfy the stronger
requirement of monotonicity, i.e.: for all $(t, u'), (t', u) \in M$
such that $t' \preceq t$ it holds $u' \not\prec u$. The reader may
wonder why we then need the freedom offered by the weaker requirement
of cross--closedness. As an example of an important type matching
relation that is cross--closed but not monotone we mention the
standard C type--promotion rules (restricted to $\mt{int}$ and
$\mt{char}$):
\begin{align*}
  R_\ms{promote} = \{ \mt{char} \mapsto \mt{int}, \mt{char} \mapsto
  \mt{char}, \mt{int} \mapsto \mt{int} \}
\end{align*}
We need a relation such as this one to deal with the standard C
promotion rules properly. As can be seen this relation is not monotone
as $\mt{char}$ can be promoted to $\mt{int}$ \emph{or} to itself. For
space constraints we simplify the treatment of function calls, meaning
we will not apply the promotion rules in the remainder of this
paper. We just mention that, in practice, promotion rules can easily
be implemented by introducing such an additional relation
$R_\ms{promote}$ between the outer argument expression syntax node and
the inner function argument syntax node.

\subsection{Storing Ambiguous Types as Antichains}\label{sec:antichains}

The goal of the typing procedure that we will describe in this section
will be to arrive at a \concept{simple typing} as introduced in
Definition~\ref{def:syntax_graphs}. However, \emph{before} this goal
is reached, so \emph{during} the typing process, it may occur that we
must keep more than one alternative type as the information that is
incident on a node from various directions is being processed.

In Figure~\ref{fig:typing:fun:c} this is illustrated: because the type of
the first argument node 1 is not yet fixed (perhaps this information
is still being propagated elsewhere in the syntax graph) we have to
keep two alternative types for the signature node 3. Despite this
ambiguity in function signature, the type of the result can be known
none--the--less and is being propagated upward in the syntax graph to
node 4.

Typically, type ambiguity would be dealt with by moving to the full
powerset lattice of types, i.e. we would annotate the nodes of the
syntax graph with \emph{subsets of types} rather than individual
concrete types. The downside of this is that the typing process
becomes prohibitively expensive to perform because we must keep track
of arbitrary subsets of the set of relevant types. For this reason we
propose to move to the lattice of \concept{antichains of types}
instead.

This has the advantage of providing a useful form of
abstraction. Because we are approximating the valid typing from above
in the lattice of antichains of types it means we may, at all times,
\concept{restrict to the maximal (weakest) types}. In practice this is
efficient and the loss in precision turns out to be
modest. Overfitting of typings is automatically ruled out: typing
ambiguities are always detected.
Formally the lattice of antichains of types is defined as follows.
\begin{definition}[Antichains]\label{def:antichains}
An \concept{antichain of types} $A \subseteq T$ is a set of types that
are \concept{pairwise incomparable}, i.e.: for all $t, t' \in A$ it
holds neither $t \prec t'$ nor $t' \prec t$. For a given subset of
types $U \subseteq T$ with $U^\mt{+}$ we denote the \concept{downward
  closure} of $U$, defined as $U^\mt{+} = \{ t \in T\ |\ \exists u \in
U . t \preceq u \}$, with $\lceil U \rceil$ we denote the restriction
of $U$ to \concept{maximal elements} defined as $\lceil U \rceil = \{
u \in U\ |\ \nexists u' \in U . u \prec u' \}$. Note that $U \subseteq
T$ is an antichain iff $\lceil U \rceil = U$. With $\mc{A}[T]$ we
denote the set of antichains of types $\mc{A}[T] = \{ U \subseteq
T\ |\ \lceil U \rceil = U \}$. We define an ordering $\sqsubseteq$ on
antichains such that $A \sqsubseteq B$ iff $\forall a \in A. \exists b
\in B. a \preceq b$, we say $B$ subsumes $A$. Note that $A \sqsubseteq
B$ iff $A^\mt{+} \subseteq B^\mt{+}$. We define $A \sqcup B = \lceil A
\cup B \rceil$. Note that $(A \sqcup B)^\mt{+} = A^\mt{+} \cup
B^\mt{+}$. Finally note that $\langle \mc{A}[T], \sqsubseteq, \sqcup
\rangle$ forms a complete lattice.
\end{definition}
In Figure~\ref{fig:hierarchy:2} we show an example of an antichain of
function types together with the associated downward closed set of
types. We consider antichains an efficient, compact symbolic
representation of the downward closed set of types. For example, the
antichain annotating node 3 in Figure~\ref{fig:typing:fun:c} should be
interpreted as such: the node must be typed with some type from the
downward closed set of types spanned by its antichain as shown in
Figure~\ref{fig:hierarchy:2}. 

If for example, we would now introduce the additional information that
node 1 is of type \verb+int+ then the antichain for node 3 would
converge further to the singleton $\{ \verb+int(*)(int, int)+
\}$. This is exactly the goal of the typing procedure: to arrive at a
singleton antichain so that the node has a single well defined
type. The following definition makes this precise.

\begin{definition}[Strengthenable Typing] 
For a given syntax graph $G$ we define a \concept{strengthenable typing} as
a map $\Delta : N \to \mc{A}[T]$ assigning each node an antichain of
types. If for some $n \in N$ it holds $|\Delta(n)| > 1$ we say the
strengthenable typing is \concept{ambiguous}. If for some $n \in N$ it holds
$\Delta(n) = \nada$ we say the typing is \concept{inconsistent}. 
If $\Delta$ is consistent and non--ambiguous it may be turned into a
simple typing $\delta$ such that $\delta(n) = t$ iff $\Delta(n) = \{ t
\}$, in this case we say $\Delta$ is \emph{valid} iff $\delta$ is
valid. We define an ordering $\sqsubseteq$ on typings such that
$\Delta \sqsubseteq \Delta'$ iff for all $n \in N$ it holds $\Delta(n)
\sqsubseteq \Delta'(n)$.
\end{definition}

Note that a strengthenable typing $\Delta$ can be seen as an element
in the product lattice $\mc{A}[T]^N$. This product lattice will be the
main lattice on which we will define the bidirectional antichain
typing algorithm in Section~\ref{sec:algorithm}

\subsection{Bidirectional Antichain Typing Algorithm}\label{sec:algorithm}

\begin{algorithm}[b]
\caption{Type the given abstract syntax graph.}\label{alg:typing}
\begin{algorithmic}[1]
\REQUIRE A syntax graph: $G = \langle N, E, \gamma \rangle$, an
initial typing: $\Delta$
\STATE $\ms{Waiting} \leftarrow E$
\WHILE{$\ms{Waiting} \ne \nada$}
  \STATE $(n, m) \leftarrow \ms{selectFrom}(\ms{Waiting})$
  \STATE $(A, B) \leftarrow F_{\gamma(n, m)}(\Delta(n), \Delta(m))$
  \IF{$A \ne \Delta(n)$}
    \STATE $\Delta(n) \leftarrow A$
    \STATE $\ms{Waiting} = \ms{Waiting} \cup \{ (n', m') \in
    E\ |\ n \in ( n', m' ) \}$
  \ENDIF
 \IF{$B \ne \Delta(m)$}
    \STATE $\Delta(m) \leftarrow B$
    \STATE $\ms{Waiting} = \ms{Waiting} \cup \{ (n', m') \in E\ |\ m
    \in (n', m') \}$
  \ENDIF
  \STATE $\ms{Waiting} \leftarrow \ms{Waiting} \setminus \{ (n, m) \}$
\ENDWHILE
\end{algorithmic}
\end{algorithm}

In this section we present the bidirectional antichain typing
algorithm proper. The algorithm will work by approximating, from
above, a valid typing in the lattice of strengthenable
typings. Starting from some initial typing (which should reflect the
type declarations and type--casts provided by the user) we descend in
the lattice of strengthenable typings by propagating type information
along the edges of the syntax graph through application of a
bidirectional flow function. The following definition makes this
precise.

\begin{definition}[Antichain Typing Algorithm]\label{def:flow}
For each $\gamma \in \Gamma$ we define a \concept{bidirectional flow
  function} $F_\gamma : \mc{A}[T]^2 \to \mc{A}[T]^2$ such that for all
$A, B \in \mc{A}[T]$ it holds:
\begin{align*}
F_\gamma( A, B ) = ( \ 
& \lceil \{ a \in A^\mt{+}\ |\ \exists b \in B^\mt{+} . (a, b) \in M_\gamma
\} \rceil, \\
& \lceil \{ b \in B^\mt{+}\ |\ \exists a \in A^\mt{+} . (a, b) \in M_\gamma
\} \rceil\ )
\end{align*}
i.e. the new antichains consist of the weakest types from the left and
right downward closed sets for which there exists at least one
underlying type relation edge into the opposing downward closed set.
\end{definition}

In Section~\ref{sec:efficient} we show how to compute the
bidirectional flow function efficiently, avoiding iteration over the
Cartesian product of the downward closed sets. There we also explain
why the function cannot easily be split in two separate flow
functions.
The \concept{Bidirectional Antichain Typing Algorithm}
Algorithm~\ref{alg:typing} shows the bidirectional flow function being
applied.
Because the flow function is bidirectional the algorithm can infer the
type of the argument expressions of some operator expression based on
the required result type of the operator expression. This allows us to
deal with the situation where the type of a declared identifier must
be inferred from the required result type of the operations in which it
takes part.
To illustrate the algorithm we first look at three different
applications of its basic step: the bidirectional flow function.

\begin{figure}\center
\subfigure[Syntax
  Graph\label{fig:typing:fwd:a}]{\input{figures/typing_fwd_a.tex}}
\subfigure[Initial
  Typing\label{fig:typing:fwd:b}]{\input{figures/typing_fwd_b.tex}}
\subfigure[Next
  Typing\label{fig:typing:fwd:c}]{\input{figures/typing_fwd_c.tex}}
\caption{Example of type information flowing forward.}\label{fig:typing:fwd}
\end{figure}
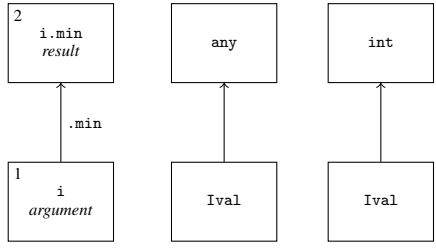
In Figure~\ref{fig:typing:fwd} we show how type information can be
propagated in a forward direction. First we show, in
Figure~\ref{fig:typing:fwd:a}, the syntax graph for the expression
\verb+i.min+. Next we show, in Figure~\ref{fig:typing:fwd:b}, an
initial typing that assigns the \verb+Ival+ type to the argument node
of the field select expression and the \verb+any+ type to the result
node. Finally we show, in Figure~\ref{fig:typing:fwd:c} the result of
applying $F_\mt{.min}$ once on the initial typing. As can be seen the
net effect of the flow function is that the type of the field
(\verb+int+) is derived from the type of the \verb+struct+
(\verb+Ival+). This constitutes one particular example where type
flows in a \emph{forward direction} over the edges of the syntax
graph.

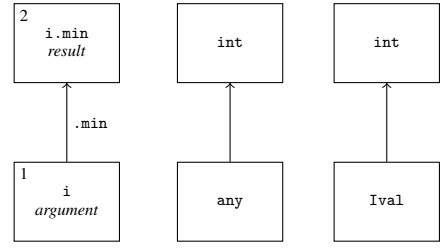
\begin{figure}[t]\center
\subfigure[Syntax Graph\label{fig:typing:bwd:a}]
  {\input{figures/typing_bwd_a.tex}}
\subfigure[Initial Typing\label{fig:typing:bwd:b}]
  {\input{figures/typing_bwd_b.tex}}
\subfigure[Next Typing\label{fig:typing:bwd:c}]
  {\input{figures/typing_bwd_c.tex}}
\caption{Example of type information flowing backward.}\label{fig:typing:bwd}
\end{figure}
In Figure~\ref{fig:typing:bwd} we show how type information can be
propagated in a backward direction. First we show, in
Figure~\ref{fig:typing:bwd:a}, the syntax graph for the expression
\verb+i.min+. Next we show, in Figure~\ref{fig:typing:bwd:b}, an
initial typing that assigns the \verb+int+ type to the result node of
the field select expression and the \verb+any+ type to the argument
node. Finally we show, in Figure~\ref{fig:typing:bwd:c} the result of
applying $F_\mt{.min}$ once on the initial typing. As can be seen the
net effect of the flow function is that the type of the \verb+struct+
(\verb+Ival+) is derived from the type of the field (\verb+int+). This
constitutes one particular example where type flows in a
\emph{backward direction} over the edges of the syntax graph.

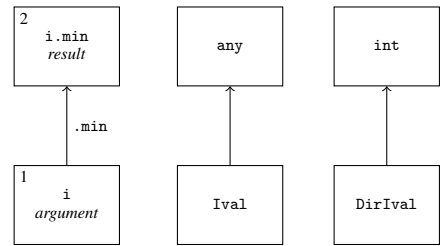
\begin{figure}[b]\center
\subfigure[Syntax Graph\label{fig:typing:bidi:a}]
  {\input{figures/typing_bidi_a.tex}}
\subfigure[Initial Typing\label{fig:typing:bidi:b}]
  {\input{figures/typing_bidi_b.tex}}
\subfigure[Next Typing\label{fig:typing:bidi:c}]
  {\input{figures/typing_bidi_c.tex}}
\caption{Example of type information flowing both ways.}\label{fig:typing1bidi}
\end{figure}
In Figure~\ref{fig:typing1bidi} we show how type information can be
propagated in a bidirectional fashion. First we show, in
Figure~\ref{fig:typing:bidi:a}, the syntax graph for the expression
\verb+i.delta+. Next we show, in Figure~\ref{fig:typing:bidi:b}, an
initial typing that assigns the \verb+any+ type to the result node of
the field select expression and the \verb+Ival+ type to the argument
node. Finally we show, in Figure~\ref{fig:typing:bidi:c} the result of
applying $F_\mt{.delta}$ once on the initial typing. As can be seen the
net effect of the flow function is that the type of the \verb+struct+
(\verb+DirIval+) and the type of the field (\verb+int+) are derived
simultaneously. This constitutes one particular example where types
flow in \emph{both directions} over the edges of the syntax graph.

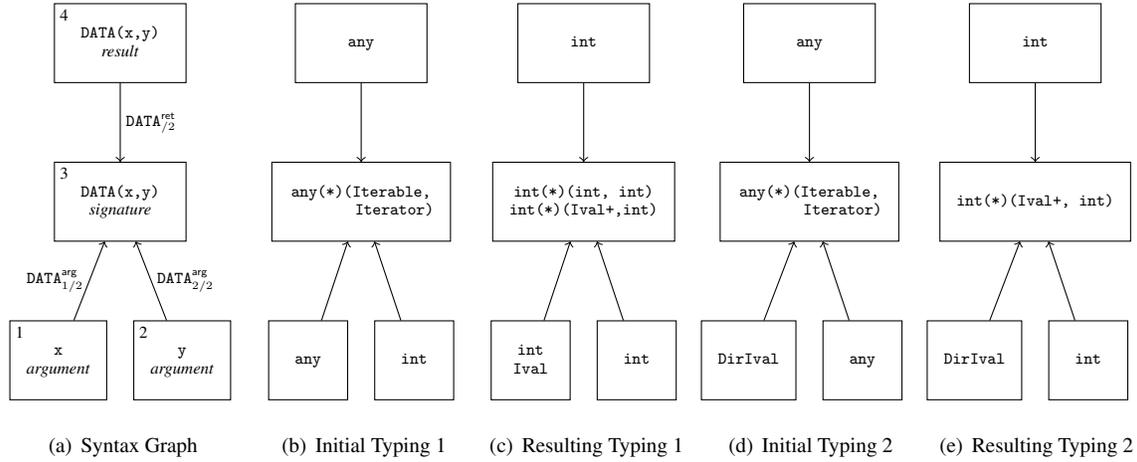
\begin{figure*}\center
\subfigure[Syntax
  Graph\label{fig:typing:fun:a}]{\input{figures/typing_fun_a.tex}}
\subfigure[Initial Typing
  1\label{fig:typing:fun:b}]{\input{figures/typing_fun_b.tex}}
\subfigure[Resulting Typing
  1\label{fig:typing:fun:c}]{\input{figures/typing_fun_c.tex}}
\subfigure[Initial Typing
  2\label{fig:typing:fun:d}]{\input{figures/typing_fun_d.tex}}
\subfigure[Resulting Typing
  2\label{fig:typing:fun:e}]{\input{figures/typing_fun_e.tex}}
\caption{Two examples of the bidirectional antichain typing algorithm
  when run to completion for two different initial strengthenable
  typings.}\label{fig:typing:fun}
\end{figure*}
Finally, in Figure~\ref{fig:typing:fun} we show two examples of how
the bidirectional antichain typing algorithm solves typing constraints
over a more complex syntax graph by repeatedly applying the
bidirectional flow function.

In Figure~\ref{fig:typing:fun:a} we show the syntax graph of the
function call expression \verb+DATA(x,y)+ from the example in
Section~\ref{sec:motivating}. Next we show, in
Figure~\ref{fig:typing:fun:b}, an initial typing that assigns
\verb+int+ to the second argument node and an uninformed type to all
the other nodes, including the signature node. Next we show, in
Figure~\ref{fig:typing:fun:b} what is the result of running algorithm
1 on this fragment of the syntax graph. In other words: we repeatedly
apply the bidirectional flow function until nothing changes
anymore. As can be seen the net the result node receives a singleton
type \verb+int+ but the signature node and the left argument node are
still ambiguous because there is not enough information to determine a
singleton type for these nodes. In absence of other information we
will have to reject the program with an ambiguous typing error in this
case.

In Figure~\ref{fig:typing:fun:d} we show an initial typing that
assigns \verb+DirIval+ to the first argument node and an uninformed
type to all the other nodes. Next we show, in
Figure~\ref{fig:typing:fun:e} what is the result of running algorithm
1 on this fragment of the syntax graph. As can be seen \emph{all} the
nodes receive a singleton type. In this case the typing converged and
we know which function body to call. The signature node 3 determines
the actual function that should be invoked, and the surrounding
parameter and return type nodes 1, 2 and 4 determine the actual
parameters and return type of the function.

If this is the first time we encounter a call to \verb+DATA+ with
these actual parameter types and return type we start the typing
procedure on a fresh copy of the syntax graph for the function body of
\verb+DATA+ strengthened using the actual argument and return types
from nodes 1, 2 and 4 respectively. If, in the process we find a
strengthening of one of the arguments and return type of \verb+DATA+
we may propagate this information back to the nodes 1, 2 and 4
respectively. This is a convenient way to handle both the \emph{inter}
as well as the \emph{intra}--procedural type flow in a uniform
way. Without the need to explicitly deal with higher order
type--parameters during the typing process. Note that this procedure
is still guaranteed to terminate because the number of types and the
number of function definitions is finite, therefore also the number of
possible function strengthenings is finite.

We keep the various typings of each function in a spanning tree. This
approach allows us to give a lot of context information when typing
errors occur. For example, a type error in \verb+DATA+:
\begin{samepage}
\begin{verbatim}
type error after call sequence:
1: int main(int, char**)
2: void print( DirIval, char* )
3: void print( DirIval )
4: int DATA( DirIval, int )
...
\end{verbatim}
\end{samepage}
Note that this is not a ``real'' call--stack as recursive calls are
collapsed over their actual signature.

\subsection{Correctness}

The correctness of Algorithm~\ref{alg:typing} is ensured by the
following two theorems.

\begin{samepage}
\begin{theorem}[Soundness]
After termination of Algorithm~\ref{alg:typing}, if $\Delta$ is
consistent and non--ambiguous then $\Delta$ is valid.
\end{theorem}
\emph{Proof}.
Assume closure is reached and $\Delta$ is unambiguous and consistent
so that we may define $\delta(n) = t$ iff $\Delta(n) = \{t\}$. We
prove that $\delta$ is valid. \emph{W.l.o.g. consider some edge} $(n,
m) \in E$ and let $\delta(n) = t$ and $\delta(m) = u$ and $F =
F_\gamma(n, m)$ and $M = M_\gamma(n, m)$. Because closure was reached
it must hold $F(\{t\}, \{u\}) = (\{t\}, \{u\})$ by definition of $F$
this implies that there must exist $t' \preceq t$ such that $(t', u)
\in M$ and there must exist $u' \preceq u$ such that $(t, u') \in
M$. By cross--closedness this implies that $(t, u) \in M$ as
required. $\Box$
\end{samepage}

\begin{samepage}
\begin{theorem}[Termination]
Algorithm~\ref{alg:typing} will always terminate.
\end{theorem}
\emph{Proof}. First observe that the bidirectional flow function
$F_\gamma$ is monotone over the lattice of typings. Next note that the
lattice $\mc{A}[T]^N$ of typings is finite, so we can go down only a
finite number of times before the waitinglist becomes empty. $\Box$
\end{samepage}

\subsection{Efficient Implementation of the Flow Function}\label{sec:efficient}

In Definition~\ref{def:flow} we introduced the bidirectional flow
function. However, this definition depends on the Cartesian product
between the downward closed sets $A^\mt{+}$ and $B^\mt{+}$ of types.

This means that a naive, direct implementation of the flow function
would need to iterate over this product in order to compute the
resulting pair of antichains. This defeats the purpose of using
antichains as a symbolic representation for their downward closed
sets. Especially for generic types the downward closed sets can become
quite large. As a particular example, the downward closed set of the
top antichain $\{ \verb+any+ \}^\mt{+}$ contains \emph{all} the
relevant types $T$.

Therefore, in this section, we give an alternative, equivalent
formulation of the bidirectional flow function which avoids this
explicit iteration over the downward closed sets. The following
definition makes this precise.

\begin{definition}[Symbolic Flow Functions]\label{def:efficient}
We lift the standard join $\sqcup$ on antichains of types as given in
Definition~\ref{def:antichains} to \emph{pairs} of antichains such
that $(A, B) \sqcup (A', B') = (A \sqcup A', B \sqcup B')$. We now
define the symbolic bidirectional flow function $\mc{F}_\gamma$ for a
given type matching relation $\gamma \in \Gamma$ such that
\begin{align*}
\mc{F}_\gamma(A, B) = \bigsqcup_{\underline{a} \in A, \underline{b} \in B}
F_\gamma(\{ \underline{a} \}, \{ \underline{b} \} )
\end{align*}
\ 
\end{definition}
This formulation avoids an explicit enumeration over the Cartesian
product of the downward closed sets $A^\mt{+}$ and $B^\mt{+}$, and
instead expresses the flow function for arbitrary antichains in terms
of the Cartesian product of $A$ and $B$ directly (note that $A$ and
$B$ are usually much smaller than $A^\mt{+}$ and $B^\mt{+}$, and,
typically even singletons). The inner call is still an application to
the old definition of $F_\gamma$ but this is a very specific case,
namely: $F_\gamma$ applied only to singleton antichains. In effect,
the fact that these antichains are singleton means they now represent
\emph{concrete} types.

The following theorem ensures the soundness of this optimization.
\begin{samepage}
\begin{theorem}
It holds $\mc{F}_\gamma(A, B) = F_\gamma(A, B)$.
\end{theorem}
\emph{Proof.} Starting from the definition of $F_\gamma$:
\begin{align*}
F_\gamma( A, &B ) = \\ 
&(\ \lceil \{ a \in A^\mt{+}\ |\ \exists b \in B^\mt{+} . (a, b) \in
M_\gamma \} \rceil, \\
& \ \ \;\lceil \{ b \in B^\mt{+}\ |\ \exists a \in A^\mt{+} . (a, b)
\in M_\gamma \} \rceil\ ) \\
F_\gamma( A, &B ) = \\
&(\ \lceil \cup_{\underline{a} \in A, \underline{b} \in B} \{ a \in \{
\underline{a} \}^\mt{+}\ |\ \exists b \in \{ \underline{b} \}^\mt{+}
. (a, b) \in M_\gamma \} \rceil, \\
& \ \ \;\lceil \cup_{\underline{a} \in A, \underline{b} \in B} \{ b
\in \{ \underline{b} \}^\mt{+}\ |\ \exists a \in \{ \underline{a}
\}^\mt{+} . (a, b) \in M_\gamma \} \rceil\ ) \\
F_\gamma( A, &B ) = \\
&(\ \sqcup_{\underline{a} \in A, \underline{b} \in B} \lceil \{ a \in
\{ \underline{a} \}^\mt{+}\ |\ \exists b \in \{ \underline{b}
\}^\mt{+} . (a, b) \in M_\gamma \} \rceil, \\
& \ \;\sqcup_{\underline{a} \in A, \underline{b} \in B} \lceil \{ b
\in \{ \underline{b} \}^\mt{+}\ |\ \exists a \in \{ \underline{a}
\}^\mt{+} . (a, b) \in M_\gamma \} \rceil\ ) \\
F_\gamma(A, &B) = \\
& \sqcup_{\underline{a} \in A, \underline{b} \in B} F(\{ \underline{a}
\}, \{ \underline{b} \} ) = \mc{F}_\gamma(A, B)
\end{align*}
$\Box$
\end{samepage}

The utility of this definition is that it reduces the flow function
for arbitrary antichains to a finite antichain join over $F_\gamma$
applied to singleton flow pairs. These singleton flow pairs can easily
be pre--computed for each relevant pair. Note that, in particular,
$F_\gamma(\{ t \}, \{ u \}) = (\{ t \}, \{ u \})$ iff $(t, u) \in
M_\gamma$. From this observation it is not hard to develop a closure
procedure that efficiently pre--computes $F_\gamma$ for all the
relevant singleton pairs that lead to non--empty resulting pairs.

In practice we will pre--compute $F_\gamma$ for all relevant singleton
pairs and place the results in a sparse lookup table for dynamic
programming. Whenever the algorithm requests an evaluation of
$F_\gamma$ on some pair $(A, B)$ of non--singleton antichains that has
not been seen before we use definition~\ref{def:efficient} to reduce
the result to a finite join over $F_\gamma$ applied to singleton pairs
(which are guaranteed to be in the lookup table). Once the result is
known we add it to the lookup table so that the next time we evaluate
$F_\gamma(A, B)$ this will come at the cost of a single
lookup. Because the bulk of the evaluations to $F_\gamma$ are highly
repetitive this greatly speeds up the typing process.

\section{Conclusion}\label{sec:conclusion}

In this paper we have presented a new approach to static typing of
generic container types written in an object oriented style. Our
contribution is twofold. First we have shown how to use an adapted
definition of a simulation pre--order to automatically infer
type--hierarchy in a structural type system that supports argument
strengthening. Second we have shown how to use an antichain based
representation for types to efficiently implement the resulting type
system.

The choice to base the present work on the C++ programming language is
mainly a pragmatic one. For the type of high performance (scientific)
software that formed the impetus for this work C(++) still constitutes
the de facto standard, also because of the huge amount of legacy code
and libraries that are available.
Nevertheless we believe the techniques outlined in this paper are more
generally applicable, in particular programming layers for other
programming languages can be similarly defined.

As future work on {\sc moot} we are planning to include dynamic type
subsumption, separate compilation and memory locality into the
programming layer.

\subsection{Acknowledgments}

We would like to thank Jean--Fran\c cois Raskin and Nicolas Maquet for
helpful suggestions, comments and inspiring discussions during the
preparation of this paper.

\bibliographystyle{plain} \bibliography{moot-paper}

\end{document}

%% file: figures/hierarchy_1.tex
\begin{tikzpicture}[scale=.7,transform shape]
\node (any) [draw,rectangle,minimum width=2cm, minimum height=.8cm]
      {{\color{white}I}\verb+any+{\color{white}I}};
\node (Iterator) at ([yshift=-1cm,xshift=-.25cm] any.south west)
      [anchor=north east,draw,rectangle,minimum width=2cm, minimum
        height=.8cm] {\verb+Iterator+};
\node (Iterable) at ([yshift=-1cm,xshift=.25cm] any.south east)
      [anchor=north west,draw,rectangle,minimum width=2cm, minimum
        height=.8cm] {\verb-Iterable-};
\node (Ival) at ([yshift=-1.2cm,xshift=2cm] Iterable.south)
      [anchor=north,draw,rectangle,minimum width=2cm, minimum
        height=.8cm] {\verb-Ival-};
\node (DirIval) at ([yshift=-1cm] Ival.south)
      [anchor=north,draw,rectangle,minimum width=2cm, minimum
        height=.8cm] {\verb+DirIval+};
\node (int) at ([yshift=-3cm,xshift=0cm] any.south)
      [anchor=north,draw,rectangle,minimum width=2cm, minimum
        height=.8cm] {\verb+int+};
\node (cadre) at ([yshift=-.75cm,xshift=1.1cm] DirIval.south east)
      [anchor=south east,minimum width=9.75cm,minimum
        height=7.25cm,rectangle,style=dotted,use as bounding box]
      {};
\draw (any) -- (Iterable);
\draw (any) -- (Iterator);
\draw (Iterable) -- (int);
\draw (Iterable) -- (Ival);
\draw (Ival) -- (DirIval);
\draw (Iterator) -- (int);
\draw [style=dashed] ([yshift=.5cm] Iterable.north) .. controls
([xshift=4cm,yshift=.5cm] Iterable.north) and ([xshift=4cm,yshift=.5cm]
Iterable.north) .. ([xshift=4cm,yshift=-5cm] Iterable.north);
\draw [style=dashed] ([yshift=.5cm] Iterable.north) .. controls
([xshift=-2cm,yshift=.5cm] Iterable.north) and ([xshift=-4cm,yshift=.5cm]
Iterable.north) .. ([xshift=-4cm,yshift=-5cm] Iterable.north);
\draw [style=dashed] ([yshift=.5cm] Ival.north) .. controls
([xshift=1.5cm,yshift=.5cm] Ival.north) and ([xshift=1.5cm,yshift=.5cm]
Ival.north) .. ([xshift=1.5cm,yshift=-3cm] Ival.north);
\draw [style=dashed] ([yshift=.5cm] Ival.north) .. controls
([xshift=-1.5cm,yshift=.5cm] Ival.north) and ([xshift=-1.5cm,yshift=.5cm]
Ival.north) .. ([xshift=-1.5cm,yshift=-3cm] Ival.north);
\end{tikzpicture}

%% file: figures/classdiag.tex
\begin{tikzpicture}[scale=.7,transform shape]
\node (ArrayList) [rectangle,minimum width=4.5cm, minimum
  height=.8cm,anchor=north east] at ([xshift=-.5cm,yshift=-1cm] any.south west)
      {\begin{minipage}{4cm}
       \verb+ArrayList+
       \end{minipage}};
\node (ArrayListElems) [rectangle,minimum width=4.5cm, minimum
  height=.8cm,anchor=north] at ([yshift=-0pt] ArrayList.south)
      {\begin{minipage}{4cm}
       \verb-ArrayListData *elems-
       \end{minipage}};
\node (ArrayListSize) [rectangle,minimum width=4.5cm, minimum
  height=.8cm,anchor=north] at ([yshift=.25cm] ArrayListElems.south)
      {\begin{minipage}{4cm}
       \verb-int size-
       \end{minipage}};
\node (ArrayListData) [rectangle,minimum width=4.5cm, minimum
  height=.8cm,anchor=west] at ([xshift=2cm]
ArrayListElems.east) {
  \begin{minipage}{4cm}
  \verb+any <<ArrayListData>>+
  \end{minipage}};
\node (IvalList) [rectangle,minimum width=4.5cm, minimum
  height=.8cm,anchor=north] at ([yshift=-1.5cm]
ArrayListSize.south) 
   {\begin{minipage}{4cm}
    \verb+IvalList+
    \end{minipage}};
\node (IvalListElems) [rectangle,minimum width=4.5cm, minimum
  height=.8cm,anchor=north] at ([yshift=-0pt] IvalList.south)
      {\begin{minipage}{4cm}
       \verb-Ival *elems-
       \end{minipage}};
\node (Ival) [rectangle,minimum width=4.5cm, minimum
  height=.8cm,anchor=west] at ([xshift=2cm]
IvalListElems.east) 
  {\begin{minipage}{4cm}
   \verb+Ival+
   \end{minipage}};
\node (IvalMin) [rectangle,minimum width=4.5cm, minimum
  height=.8cm,anchor=north] at ([yshift=-0pt] Ival.south)
      {\begin{minipage}{4cm}
       \verb-int min-
       \end{minipage}};
\node (IvalMax) [rectangle,minimum width=4.5cm, minimum
  height=.8cm,anchor=north] at ([yshift=.25cm] IvalMin.south)
      {\begin{minipage}{4cm}
       \verb-int max-
       \end{minipage}};
\node (DirIvalList) [rectangle,minimum width=4.5cm, minimum
  height=.8cm,anchor=north] at ([yshift=-1.5cm]
IvalListElems.south) {
    \begin{minipage}{4cm}
    \verb+DirIvalList+
    \end{minipage}};
\node (DirIvalListElems) [rectangle,minimum width=4.5cm, minimum
  height=.8cm,anchor=north] at ([yshift=-0pt] DirIvalList.south)
      {\begin{minipage}{4cm}
       \verb-DirIval *elems-
       \end{minipage}};
\node (DirIval) [rectangle,minimum width=4.5cm, minimum
  height=.8cm,anchor=west] at ([xshift=2cm]
DirIvalListElems.east) 
   {\begin{minipage}{4cm}
    \verb+DirIval+
    \end{minipage}};
\node (DirIvalDelta) [rectangle,minimum width=4.5cm, minimum
  height=.8cm,anchor=north] at ([yshift=-0pt] DirIval.south)
      {\begin{minipage}{4cm}
       \verb-int delta-
       \end{minipage}};
\node (cadre) [rectangle,minimum width=12cm, minimum height=10.5cm,
  anchor=north west] at ([xshift=-.5cm,yshift=1cm] ArrayList.north
west) {};
\draw (ArrayList.north west) -- (ArrayList.north east) --
(ArrayListSize.south east) -- (ArrayListSize.south west) --
(ArrayList.north west);
\draw (ArrayListData.north west) -- (ArrayListData.north east) --
(ArrayListData.south east) -- (ArrayListData.south west) --
(ArrayListData.north west);
\draw (ArrayList.south west) -- (ArrayList.south east);
\draw (Ival.north west) -- (Ival.north east) --
(IvalMax.south east) -- (IvalMax.south west) --
(Ival.north west);
\draw (Ival.south west) -- (Ival.south east);
\draw (IvalList.north west) -- (IvalList.north east) --
(IvalListElems.south east) -- (IvalListElems.south west) --
(IvalList.north west);
\draw (IvalList.south west) -- (IvalList.south east);
\draw (DirIvalList.north west) -- (DirIvalList.north east) --
(DirIvalListElems.south east) -- (DirIvalListElems.south west) --
(DirIvalList.north west);
\draw (DirIvalList.south west) -- (DirIvalList.south east);
\draw (DirIval.north west) -- (DirIval.north east) --
(DirIvalDelta.south east) -- (DirIvalDelta.south west) --
(DirIval.north west);
\draw (DirIval.south west) -- (DirIval.south east);
\draw [->,>=open triangle 45] (ArrayList.north) -- ([yshift=.5cm]
ArrayList.north) node (p1) {} -- (p1 -| ArrayListData.north) --
(ArrayListData.north);
\draw [->,>=open triangle 45] (IvalList.north) --
(ArrayListSize.south);
\draw [->,>=open triangle 45] (Ival.north) --
(ArrayListData.south);
\draw [->,>=open triangle 45] (DirIvalList.north) --
(IvalListElems.south);
\draw [->,>=open triangle 45] (DirIval.north) --
(IvalMax.south);
\draw [->,>=open diamond] (ArrayListData.west) node [xshift=-4mm,above] {0..*}
-- (ArrayListElems.east) node [xshift=4mm,above] {};
\draw [->,>=open diamond] (Ival.west) node [xshift=-4mm,above] {0..*}
-- (IvalListElems.east) node [xshift=4mm,above] {};
\draw [->,>=open diamond] (DirIval.west) node [xshift=-4mm,above] {0..*}
-- (DirIvalListElems.east) node [xshift=4mm,above] {};
\end{tikzpicture}

%% file: figures/late.tex
\begin{tikzpicture}[scale=.7,transform shape]
\node (source) 
      [minimum width=3cm,minimum height=1.5cm,draw,rectangle]
      {\begin{tabular}{c}C++ template\\source\end{tabular}};
\node (instantiate) at ([yshift=-.5cm] source.south)
      [anchor=north,minimum width=3cm,minimum
        height=1.5cm,draw,rectangle, rounded corners]
      {\begin{tabular}{c}template\\instantiation\end{tabular}};
\node (ssource) at ([yshift=-.5cm] instantiate.south)
      [anchor=north,minimum width=3cm,minimum
        height=1.5cm,draw,rectangle]
      {\begin{tabular}{c}C++\\code\end{tabular}};
\node (typecheck) at ([yshift=-.5cm] ssource.south)
      [anchor=north,minimum width=3cm,minimum
        height=1.5cm,draw,rectangle, rounded corners]
      {\begin{tabular}{c}C++\\typechecker\end{tabular}};
\node (late_warning) at ([xshift=1cm] typecheck.east)
      [anchor=west,minimum width=4.5cm,minimum
        height=1.5cm,style=dotted,draw,rectangle,rounded corners]
      {\begin{tabular}{c}C++\\errors\end{tabular}};
\draw [->] (source) -- (instantiate);
\draw [->] (instantiate) -- (ssource);
\draw [->] (ssource) -- (typecheck);
\draw [->] (typecheck.south) -- ([yshift=-.5cm] typecheck.south) node
      {$\vdots$};
\draw [->,style=dotted] (typecheck) -- (late_warning);
\end{tikzpicture}

%% file: figures/early.tex
\begin{tikzpicture}[scale=.7,transform shape]
\node (source) 
      [minimum width=3cm,minimum height=1.5cm,draw,rectangle]
      {\begin{tabular}{c}{\sc moot}\\source\end{tabular}};
\node (instantiate) at ([yshift=-.5cm] source.south)
      [anchor=north,minimum width=3cm,minimum
        height=1.5cm,draw,rectangle, rounded corners]
      {\begin{tabular}{c}type parameter\\instantiation\end{tabular}};
\node (typestr) at ([yshift=-.5cm] instantiate.south)
      [anchor=north,minimum width=3cm,minimum
        height=1.5cm,draw,rectangle, rounded corners]
      {\begin{tabular}{c}function type\\strengthening\end{tabular}};
\node (early_warning) at ([xshift=1cm] instantiate.east)
      [anchor=west,minimum width=4.5cm,minimum
        height=1.5cm,style=dotted,draw,rectangle,rounded corners]
      {\begin{tabular}{c}type parameter\\instantiation errors\end{tabular}};
\node (early_warning2) at ([xshift=1cm] typestr.east)
      [anchor=west,minimum width=4.5cm,minimum
        height=1.5cm,style=dotted,draw,rectangle,rounded corners]
      {\begin{tabular}{c}ambiguous/inconsistent typing\\errors\end{tabular}};
\node (ssource) at ([yshift=-.5cm] typestr.south)
      [anchor=north,minimum width=3cm,minimum
        height=1.5cm,draw,rectangle]
      {\begin{tabular}{c}C++\\code\end{tabular}};
\node (typecheck) at ([yshift=-.5cm] ssource.south)
      [anchor=north,minimum width=3cm,minimum
        height=1.5cm,draw,rectangle, rounded corners]
      {\begin{tabular}{c}C++\\typechecker\end{tabular}};
\node (late_warning) at ([xshift=1cm] typecheck.east)
      [anchor=west,minimum width=4.5cm,minimum
        height=1.5cm,style=dotted,draw,rectangle,rounded corners]
      {\begin{tabular}{c}C++\\errors\end{tabular}};
\draw [->] (source) -- (instantiate);
\draw [->] (instantiate) -- (typestr);
\draw [->] (typestr) -- (ssource);
\draw [->] (ssource) -- (typecheck);
\draw [->] (typecheck.south) -- ([yshift=-.5cm] typecheck.south) node
      {$\vdots$};
\draw [->,style=dotted] (instantiate) -- (early_warning);
\draw [->,style=dotted] (typestr) -- (early_warning2);
\draw [->,style=dotted] (typecheck) -- (late_warning);
\end{tikzpicture}

%% file: figures/hierarchy_2.tex
\begin{tikzpicture}[scale=.8,transform shape]
\node (any) [draw,rectangle,minimum width=5cm, minimum height=.8cm]
      {{\color{white}I}\verb+any(*)(Iterable, Iterator)+{\color{white}I}};
\node (int) at ([yshift=-1.5cm,xshift=+2cm] any.south west)
      [anchor=north east,draw,rectangle,minimum width=3.5cm, minimum
        height=.8cm] {\verb+int(*)(int, int)+};
\node (ival) at ([yshift=-1.5cm,xshift=-2cm] any.south east)
      [anchor=north west,draw,rectangle,minimum width=3.5cm, minimum
        height=.8cm] {\verb-int(*)(Ival+, int)-};
\node (dirival) at ([yshift=-1.5cm,xshift=0cm] ival.south)
      [anchor=north,draw,rectangle,minimum width=4.5cm, minimum
        height=.8cm] {\verb-int(*)(DirIval+, int)-};
\node (cadre) at ([yshift=-1cm,xshift=1cm] dirival.south east)
      [anchor=south east,minimum width=10.5cm,minimum
        height=6.5cm,rectangle,style=dotted,use as bounding box]
      {};
\draw (any) -- (int);
\draw (any) -- (ival);
\draw (ival) -- (dirival);
\begin{scope}
\clip (cadre.north west) -- (cadre.north east) -- (cadre.south east)
-- (cadre.south west);
\draw [style=dashed] ([xshift=-.5cm,yshift=-.5cm] dirival.south west)
.. controls ([xshift=-.5cm,yshift=+1cm] dirival.south west) and
([xshift=-2cm,yshift=-1.5cm] int.north west)
.. ([xshift=-.5cm,yshift=+.5cm] int.north west) .. controls
([xshift=1cm,yshift=+1.5cm] int.north west) and
([xshift=-1cm,yshift=+2.75cm] dirival.north west) .. ([yshift=+2.75cm]
dirival.north west) .. controls ([xshift=1cm,yshift=+2.75cm]
dirival.north west) and ([xshift=-1cm,yshift=+1.5cm] ival.north east) ..
([xshift=.5cm,yshift=+.5cm] ival.north east) .. controls
([xshift=2cm,yshift=-.5cm] ival.north east) and
([xshift=.5cm,yshift=+1cm] dirival.south east) ..
([xshift=.5cm,yshift=-.5cm] dirival.south east);
\end{scope}
\end{tikzpicture}

%% file: figures/typing_fwd_a.tex
\begin{tikzpicture}[scale=.7,transform shape]
\node (imin) [minimum width=2cm,minimum height=1.5cm,draw,rectangle]
      {\begin{tabular}{c}{\tt i.min}\\\emph{result}\end{tabular}};
\node (iminid) at (imin.north west) [anchor=north west] {2};
\node (i) at ([yshift=-1.5cm] imin.south) [anchor=north,minimum
  width=2cm,minimum height=1.5cm,draw,rectangle]
      {\begin{tabular}{c}{\tt i}\\\emph{argument}\end{tabular}};
\node (iid) at (i.north west) [anchor=north west] {1};
\node (cadre) at ([xshift=0cm,yshift=-.5cm] i.south)
      [anchor=south,rectangle,minimum height=5.5cm, minimum
        width=3cm,style=dotted,use as bounding box] {};
  \draw [<-] (imin) -- node [right] {$\mt{.min}$} (i);
\end{tikzpicture}

%% file: figures/typing_fwd_b.tex
\begin{tikzpicture}[scale=.7,transform shape]
\node (imin) [minimum width=2cm,minimum height=1.5cm,draw,rectangle]
      {\begin{tabular}{c}{\tt any}\end{tabular}};
\node (i) at ([yshift=-1.5cm] imin.south) [anchor=north,minimum
  width=2cm,minimum height=1.5cm,draw,rectangle]
      {\begin{tabular}{c}{\tt Ival}\end{tabular}};
\node (cadre) at ([xshift=0cm,yshift=-.5cm] i.south)
      [anchor=south,rectangle,minimum height=5.5cm, minimum
        width=2.75cm,style=dotted,use as bounding box] {};
  \draw [<-] (imin) -- (i);
\end{tikzpicture}

%% file: figures/typing_fwd_c.tex
\begin{tikzpicture}[scale=.7,transform shape]
\node (imin) [minimum width=2cm,minimum height=1.5cm,draw,rectangle]
      {\begin{tabular}{c}{\tt int}\end{tabular}};
\node (i) at ([yshift=-1.5cm] imin.south) [anchor=north,minimum
  width=2cm,minimum height=1.5cm,draw,rectangle]
      {\begin{tabular}{c}{\tt Ival}\end{tabular}};
\node (cadre) at ([xshift=0cm,yshift=-.5cm] i.south)
      [anchor=south,rectangle,minimum height=5.5cm, minimum
        width=2.75cm,style=dotted,use as bounding box] {};
  \draw [<-] (imin) -- (i);
\end{tikzpicture}

%% file: figures/typing_bwd_a.tex
\begin{tikzpicture}[scale=.7,transform shape]
\node (imin) [minimum width=2cm,minimum height=1.5cm,draw,rectangle]
      {\begin{tabular}{c}{\tt i.min}\\\emph{result}\end{tabular}};
\node (iminid) at (imin.north west) [anchor=north west] {2};
\node (i) at ([yshift=-1.5cm] imin.south) [anchor=north,minimum
  width=2cm,minimum height=1.5cm,draw,rectangle]
      {\begin{tabular}{c}{\tt i}\\\emph{argument}\end{tabular}};
\node (iid) at (i.north west) [anchor=north west] {1};
\node (cadre) at ([xshift=0cm,yshift=-.5cm] i.south)
      [anchor=south,rectangle,minimum height=5.5cm, minimum
        width=3cm,style=dotted,use as bounding box] {};
  \draw [<-] (imin) -- node [right] {$\mt{.min}$} (i);
\end{tikzpicture}

%% file: figures/typing_bwd_b.tex
\begin{tikzpicture}[scale=.7,transform shape]
\node (imin) [minimum width=2cm,minimum height=1.5cm,draw,rectangle]
      {\begin{tabular}{c}{\tt int}\end{tabular}};
\node (i) at ([yshift=-1.5cm] imin.south) [anchor=north,minimum
  width=2cm,minimum height=1.5cm,draw,rectangle]
      {\begin{tabular}{c}{\tt any}\end{tabular}};
\node (cadre) at ([xshift=0cm,yshift=-.5cm] i.south)
      [anchor=south,rectangle,minimum height=5.5cm, minimum
        width=2.75cm,style=dotted,use as bounding box] {};
  \draw [<-] (imin) -- (i);
\end{tikzpicture}

%% file: figures/typing_bwd_c.tex
\begin{tikzpicture}[scale=.7,transform shape]
\node (imin) [minimum width=2cm,minimum height=1.5cm,draw,rectangle]
      {\begin{tabular}{c}{\tt int}\end{tabular}};
\node (i) at ([yshift=-1.5cm] imin.south) [anchor=north,minimum
  width=2cm,minimum height=1.5cm,draw,rectangle]
      {\begin{tabular}{c}{\tt Ival}\end{tabular}};
\node (cadre) at ([xshift=0cm,yshift=-.5cm] i.south)
      [anchor=south,rectangle,minimum height=5.5cm, minimum
        width=2.75cm,style=dotted,use as bounding box] {};
  \draw [<-] (imin) -- (i);
\end{tikzpicture}

%% file: figures/typing_bidi_a.tex
\begin{tikzpicture}[scale=.7,transform shape]
\node (imin) [minimum width=2cm,minimum height=1.5cm,draw,rectangle]
      {\begin{tabular}{c}{\tt i.min}\\\emph{result}\end{tabular}};
\node (iminid) at (imin.north west) [anchor=north west] {2};
\node (i) at ([yshift=-1.5cm] imin.south) [anchor=north,minimum
  width=2cm,minimum height=1.5cm,draw,rectangle]
      {\begin{tabular}{c}{\tt i}\\\emph{argument}\end{tabular}};
\node (iid) at (i.north west) [anchor=north west] {1};
\node (cadre) at ([xshift=0cm,yshift=-.5cm] i.south)
      [anchor=south,rectangle,minimum height=5.5cm, minimum
        width=3cm,style=dotted,use as bounding box] {};
\draw [<-] (imin) -- node [right] {$\mt{.min}$} (i);
\end{tikzpicture}

%% file: figures/typing_bidi_b.tex
\begin{tikzpicture}[scale=.7,transform shape]
\node (imin) [minimum width=2cm,minimum height=1.5cm,draw,rectangle]
      {\begin{tabular}{c}{\tt any}\end{tabular}};
\node (i) at ([yshift=-1.5cm] imin.south) [anchor=north,minimum
  width=2cm,minimum height=1.5cm,draw,rectangle]
      {\begin{tabular}{c}{\tt Ival}\end{tabular}};
\node (cadre) at ([xshift=0cm,yshift=-.5cm] i.south)
      [anchor=south,rectangle,minimum height=5.5cm, minimum
        width=2.75cm,style=dotted,use as bounding box] {};
  \draw [<-] (imin) -- (i);
\end{tikzpicture}

%% file: figures/typing_bidi_c.tex
\begin{tikzpicture}[scale=.7,transform shape]
\node (imin) [minimum width=2cm,minimum height=1.5cm,draw,rectangle]
      {\begin{tabular}{c}{\tt int}\end{tabular}};
\node (i) at ([yshift=-1.5cm] imin.south) [anchor=north,minimum
  width=2cm,minimum height=1.5cm,draw,rectangle]
      {\begin{tabular}{c}{\tt DirIval}\end{tabular}};
\node (cadre) at ([xshift=0cm,yshift=-.5cm] i.south)
      [anchor=south,rectangle,minimum height=5.5cm, minimum
        width=2.75cm,style=dotted,use as bounding box] {};
  \draw [<-] (imin) -- (i);
\end{tikzpicture}

%% file: figures/typing_fun_a.tex
\begin{tikzpicture}[scale=.7,transform shape]
\node (result) [minimum width=2.5cm,minimum height=1.5cm,draw,rectangle]
      {\begin{tabular}{c}{\tt DATA(x,y)}\\\emph{result}\end{tabular}};
%
\node (resultid) at (result.north west) [anchor=north west] {4};
\node (signature) at ([yshift=-1.5cm] result.south) [anchor=north,minimum
  width=2.5cm,minimum height=1.5cm,draw,rectangle]
      {\begin{tabular}{c}{\tt DATA(x,y)}\\\emph{signature}\end{tabular}};
%
\node (signatureid) at (signature.north west) [anchor=north west] {3};
\node (arg1) at ([yshift=-1.5cm,xshift=-.25cm] signature.south)
      [anchor=north east,minimum width=1.5cm, minimum height=1.5cm,draw,
        rectangle] {\begin{tabular}{c}{\tt
            x}\\\emph{argument}\end{tabular}};
%
\node (arg1id) at (arg1.north west) [anchor=north west] {1};
\node (arg2) at ([yshift=-1.5cm,xshift=+.25cm] signature.south)
      [anchor=north west,minimum width=1.5cm,minimum
        height=1.5cm,draw,rectangle] {\begin{tabular}{c}{\tt
            y}\\\emph{argument}\end{tabular}};
%
\node (arg2id) at (arg2.north west) [anchor=north west] {2};
\node (cadre) at (signature) [anchor=center,minimum width=4.7cm, minimum
  height=8.5cm,rectangle,style=dotted,use as bounding box] {};
\draw [<-] (signature) -- node [right]
      {${\mt{DATA}^\ms{ret}_{/2}}$} (result);
\draw [<-] (signature) -- node [left]
      {$\mt{DATA}^\ms{arg}_{1/2}$\rem{ }} (arg1);
\draw [<-] (signature) -- node [right]
      {$\mt{DATA}^\ms{arg}_{2/2}$\rem{ }} (arg2);
\end{tikzpicture}

%% file: figures/typing_fun_b.tex
\begin{tikzpicture}[scale=.7,transform shape]
\node (result) [minimum width=2.5cm,minimum height=1.5cm,draw,rectangle]
      {\begin{tabular}{c}{\tt any}\end{tabular}};
%
%
\node (signature) at ([yshift=-1.5cm] result.south)
      [anchor=north,minimum width=2.5cm,minimum
        height=1.5cm,draw,rectangle] {\begin{tabular}{c}{\tt
            any(*)(Iterable,}\\$\qquad\qquad${\tt Iterator)}\end{tabular}};
%
%
\node (arg1) at ([yshift=-1.5cm,xshift=-.25cm] signature.south)
      [anchor=north east,minimum width=1.5cm, minimum height=1.5cm,draw,
        rectangle] {\begin{tabular}{c}{\tt any}\end{tabular}};
%
%
\node (arg2) at ([yshift=-1.5cm,xshift=+.25cm] signature.south)
      [anchor=north west,minimum width=1.5cm,minimum
        height=1.5cm,draw,rectangle] {\begin{tabular}{c}{\tt
            int}\end{tabular}};
%
%
\node (cadre) at (signature) [anchor=center,minimum width=4cm, minimum
  height=8.5cm,rectangle,style=dotted,use as bounding box] {};
\draw [<-] (signature) -- (result);
\draw [<-] (signature) -- (arg1);
\draw [<-] (signature) -- (arg2);
\end{tikzpicture}

%% file: figures/typing_fun_c.tex
\begin{tikzpicture}[scale=.7,transform shape]
\node (result) [minimum width=2.5cm,minimum height=1.5cm,draw,rectangle]
      {\begin{tabular}{c}{\tt int}\end{tabular}};
%
%
\node (signature) at ([yshift=-1.5cm] result.south) [anchor=north,minimum
  width=2.5cm,minimum height=1.5cm,draw,rectangle]
      {\begin{tabular}{c}{\tt int(*)(int, int)}\\{\tt int(*)(Ival+,int)}\end{tabular}};
%
%
\node (arg1) at ([yshift=-1.5cm,xshift=-.25cm] signature.south)
      [anchor=north east,minimum width=1.5cm, minimum height=1.5cm,draw,
        rectangle] {\begin{tabular}{c}{\tt int}\\{\tt Ival}\end{tabular}};
%
%
\node (arg2) at ([yshift=-1.5cm,xshift=+.25cm] signature.south)
      [anchor=north west,minimum width=1.5cm,minimum
        height=1.5cm,draw,rectangle] {\begin{tabular}{c}{\tt
            int}\end{tabular}};
%
%
\node (cadre) at (signature) [anchor=center,minimum width=4cm, minimum
  height=8.5cm,rectangle,style=dotted,use as bounding box] {};
\draw [<-] (signature) -- (result);
\draw [<-] (signature) -- (arg1);
\draw [<-] (signature) -- (arg2);
\end{tikzpicture}

%% file: figures/typing_fun_d.tex
\begin{tikzpicture}[scale=.7,transform shape]
\node (result) [minimum width=2.5cm,minimum height=1.5cm,draw,rectangle]
      {\begin{tabular}{c}{\tt any}\end{tabular}};
%
%
\node (signature) at ([yshift=-1.5cm] result.south)
      [anchor=north,minimum width=2.5cm,minimum
        height=1.5cm,draw,rectangle] {\begin{tabular}{c}{\tt
            any(*)(Iterable,}\\$\qquad\qquad${\tt Iterator)}\end{tabular}};
%
%
\node (arg1) at ([yshift=-1.5cm,xshift=-.25cm] signature.south)
      [anchor=north east,minimum width=1.5cm, minimum height=1.5cm,draw,
        rectangle] {\begin{tabular}{c}{\tt DirIval}\end{tabular}};
%
%
\node (arg2) at ([yshift=-1.5cm,xshift=+.25cm] signature.south)
      [anchor=north west,minimum width=1.5cm,minimum
        height=1.5cm,draw,rectangle] {\begin{tabular}{c}{\tt
            any}\end{tabular}};
%
%
\node (cadre) at (signature) [anchor=center,minimum width=4cm, minimum
  height=8.5cm,rectangle,style=dotted,use as bounding box] {};
\draw [<-] (signature) -- (result);
\draw [<-] (signature) -- (arg1);
\draw [<-] (signature) -- (arg2);
\end{tikzpicture}

%% file: figures/typing_fun_e.tex
\begin{tikzpicture}[scale=.7,transform shape]
\node (result) [minimum width=2.5cm,minimum height=1.5cm,draw,rectangle]
      {\begin{tabular}{c}{\tt int}\end{tabular}};
%
%
\node (signature) at ([yshift=-1.5cm] result.south)
      [anchor=north,minimum width=2.5cm,minimum
        height=1.5cm,draw,rectangle] {\begin{tabular}{c}{\tt
            int(*)(Ival+, int)}\end{tabular}};
%
%
\node (arg1) at ([yshift=-1.5cm,xshift=-.25cm] signature.south)
      [anchor=north east,minimum width=1.5cm, minimum height=1.5cm,draw,
        rectangle] {\begin{tabular}{c}{\tt DirIval}\end{tabular}};
%
%
\node (arg2) at ([yshift=-1.5cm,xshift=+.25cm] signature.south)
      [anchor=north west,minimum width=1.5cm,minimum
        height=1.5cm,draw,rectangle] {\begin{tabular}{c}{\tt
            int}\end{tabular}};
%
%
\node (cadre) at (signature) [anchor=center,minimum width=4cm, minimum
  height=8.5cm,rectangle,style=dotted,use as bounding box] {};
\draw [<-] (signature) -- (result);
\draw [<-] (signature) -- (arg1);
\draw [<-] (signature) -- (arg2);
\end{tikzpicture}